\pdfoutput=1
\documentclass[a4paper,11pt]{article}

\usepackage{jheppub} 

\usepackage[T1]{fontenc} 
\usepackage{amsmath, amssymb, bm}
\usepackage{braket}
\usepackage{adjustbox}
\usepackage{latexsym}
\usepackage{wrapfig}
\usepackage{tabularx}
\usepackage{tikz}
\usetikzlibrary{calc,decorations.pathmorphing,decorations.markings}
\usepackage{tikz-feynhand}
\newcommand{\nn}{\nonumber \\}

\newcommand{\D}{{\rm d}}

\newcommand{\Amp}{\mathcal{M}}
\newcommand{\Disc}{{\rm Disc}}
\newcommand{\re}{\mathrm{Re}}
\newcommand{\im}{\mathrm{Im}}
\newcommand{\res}[1]{\mathop{\mathrm{Res}}_{#1} }

\newcommand{\tikzLine}[1]{
\begin{tikzpicture}[baseline=-2]
\foreach \x in {#1}
\draw (0.25,\x) -- (-0.25,\x);
\end{tikzpicture}}

\newcommand{\tikzAmp}[2]{
\begin{tikzpicture}[baseline=-2]
\foreach \x in {#1}
\draw (0.6,\x) -- (-0.6,\x);
\filldraw [fill=white] (0,0) circle [radius=4mm] ;
\node at (0,0) {$#2$} ;
\end{tikzpicture}}

\newcommand{\tikzABA}[3]{
\begin{tikzpicture}[baseline=-2]
\foreach \x in {#1}
\draw (-1,\x) -- (-0.5,\x);
\foreach \x in {#1}
\draw (1,\x) -- (0.5,\x);
\draw[line width=5pt] (-0.5, 0) -- (0.5,0); 
\filldraw [fill=white] (-0.5,0) circle [radius=4mm] ;
\node (O1) at (-0.5,0) {$#2$} ;
\filldraw [fill=white] (0.5,0) circle [radius=4mm] ;
\node (O2) at (0.5,0) {$#3$} ;
\end{tikzpicture}}

\newcommand{\tikzAA}[4]{
\begin{tikzpicture}[baseline=-2]
\foreach \x in {#1}
\draw (-1,\x) -- (-0.5,\x);
\foreach \x in {#1}
\draw (1,\x) -- (0.5,\x);
\foreach \y in {#2}
\draw (0.5,\y) -- (-0.5,\y);
\filldraw [fill=white] (-0.5,0) circle [radius=4mm] ;
\node at (-0.5,0) {$#3$} ;
\filldraw [fill=white] (0.5,0) circle [radius=4mm] ;
\node at (0.5,0) {$#4$} ;
\end{tikzpicture}}

\newcommand{\tikzABsdis}[5]{
\begin{tikzpicture}[baseline=-2]
\foreach \x in {#2}
\draw (0,\x) -- ($(#1*-0.4 ,\x)$);
\foreach \y in {#3}
\draw (-0.4, \y) -- (0.4, \y);
\foreach \z in {#4}
\draw[line width=5pt] (0,\z) -- ($(#1*0.4,\z)$);
\foreach \z in {#4}
\filldraw [fill=white] (0, \z) circle [radius=2mm] ;
\foreach \z in {#4}
\node at (0, \z) {$#5$} ;
\end{tikzpicture}
}

\newcommand{\tikzAB}[3]{
\begin{tikzpicture}[baseline=-2]
\foreach \x in {#2}
\draw ($(#1*-0.6,\x)$) -- (0,\x);
\draw[line width=5pt] ($(#1*0.6,0)$) -- (0,0);
\filldraw [fill=white] (0,0) circle [radius=4mm] ;
\node at (0,0) {$#3$} ;
\end{tikzpicture}
}

\newcommand{\tikzAAcross}[2]{
\begin{tikzpicture}[baseline=-2]
\draw (-0.65,0.2) -- (0.65,0.2);
\draw (-0.65,0) -- (-0.3,0);
\draw (0.65,0) -- (0.3,0);
\draw (-0.65,-0.2) -- (0.65,-0.2);
\draw (-0.3,0.1) -- (0.3,-0.1);
\filldraw [fill=white] (-0.3,0.1) circle [radius=2mm] ;
\node at (-0.3,0.1) {$#1$} ;
\filldraw [fill=white] (0.3,-0.1) circle [radius=2mm] ;
\node at (0.3,-0.1) {$#2$} ;
\end{tikzpicture}
}

\newcommand{\tikzAAs}[3]{
\begin{tikzpicture}[baseline=-2]
\foreach \x in {0.2,0,-0.2}
\draw (#1,\x) -- ($(#1*-0.6,\x)$);
\filldraw [fill=white] ($(#1*0.5,0)$) circle [radius=4mm] ;
\node at ($(#1*0.5,0)$) {$#2$} ;
\filldraw [fill=white] ($(#1*-0.3,0.1)$) circle [radius=2mm] ;
\node at ($(#1*-0.3,0.1)$) {$#3$} ;
\end{tikzpicture}
}

\newcommand{\tikzABABA}[3]{
\begin{tikzpicture}[baseline=-2]
\foreach \x in {0.2, 0, -0.2}
\draw (-1.5,\x) -- (-1, \x);
\foreach \y in {0.2, 0, -0.2}
\draw (1.5,\y) -- (1,\y);
\draw[line width=5pt] (-1, 0) -- (1,0); 
\filldraw [fill=white] (-1,0) circle [radius=4mm] ;
\node at (-1,0) {$#1$} ;
\filldraw [fill=white] (0,0) circle [radius=4mm] ;
\node at (0,0) {$#2$} ;
\filldraw [fill=white] (1,0) circle [radius=4mm] ;
\node at (1,0) {$#3$} ;
\end{tikzpicture}
}

\newcommand{\tikzAsAAs}[3]{
\begin{tikzpicture}[baseline=-2]
\draw (1.6,0.2) -- (-0.6,0.2);
\draw (1.6,0) -- (-0.6,0);
\draw (1.6,-0.2) -- (-0.6,-0.2);
\filldraw [fill=white] (0.5,0) circle [radius=4mm] ;
\node at (0.5,0) {$#2$} ;
\filldraw [fill=white] (-0.25,0.1) circle [radius=2mm] ;
\node at (-0.25,0.1) {$#1$} ;
\filldraw [fill=white] (1.25,0.1) circle [radius=2mm] ;
\node at (1.25,0.1) {$#3$} ;
\end{tikzpicture}
}

\newcommand{\tikzTriangle}[3]{
\begin{tikzpicture}[baseline=-2]
\draw (0.9,-0.2) -- (-0.9, -0.2) ;
\draw (0.9,0.2) -- (-0.9,0.2);
\draw (0.9,0) -- (0.6,0);
\draw (-0.9,0) -- (-0.6,0);
\draw (-0.6,0.1) -- (0,-0.1);
\draw (0.6,0.1) -- (0,-0.1);
\filldraw [fill=white] (-0.6,0.1) circle [radius=2mm] ;
\node at (-0.6,0.1) {$#1$} ;
\filldraw [fill=white] (0,-0.1) circle [radius=2mm] ;
\node at (0,-0.1) {$#2$} ;
\filldraw [fill=white] (0.6,0.1) circle [radius=2mm] ;
\node at (0.6,0.1) {$#3$} ;
\end{tikzpicture}
}

\newcommand{\tikzAsub}[3]{
\begin{tikzpicture}[baseline=-2]
\draw (0.6,0.2) -- (-0.6,0.2);
\draw (0.6,0) -- (-0.6,0);
\draw (0.6,-0.2) -- (-0.6,-0.2);
\filldraw [fill=white] (0,0) circle [radius=4mm] ;
\draw ($(0,0)+(70:0.4)$) to ($(0,0)+(290:0.4)$) ;
\draw ($(0,0)+(110:0.4)$) to ($(0,0)+(250:0.4)$) ;
\node at (0,0) {\scriptsize $#2$} ;
\node at (0.28,0) {\scriptsize $#3$} ;
\node at (-0.28,0) {\scriptsize $#1$} ;
\end{tikzpicture}
}

\newcommand{\Athree}[1]{
\begin{tikzpicture}[baseline=-2]
\draw (0.6,0.2) -- (-0.6,0.2);
\draw (0.6,0) -- (-0.6,0);
\draw (0.6,-0.2) -- (-0.6,-0.2);
\filldraw [fill=white] (0,0) circle [radius=4mm] ;
\node (O) at (0,0) {$#1$} ;
\end{tikzpicture}
}

\newcommand{\AmixL}{
\begin{tikzpicture}[baseline=-2]
\begin{feynhand}
\draw (0.7,-0.15) -- (-0.7,-0.15);
\propag[boson] (-0.7,0.15) -- (0, 0.15) ;
\propag[boson] (0.7,0.15) -- (0, 0.15) ;
\filldraw [fill=white] (0,0) circle [radius=4mm] ;
\draw ($(0,0)+(90:0.4)$) to [out = -90, in = 170] ($(0,0)+(-10:0.4)$) ;
\node (O1) at (0.2,0.2) {$-$} ;
\node (O2) at (-0.05,-0.05) {$+$} ;
\end{feynhand}
\end{tikzpicture}}

\newcommand{\AmixR}{
\begin{tikzpicture}[baseline=-2]
\begin{feynhand}
\draw (0.7,-0.15) -- (-0.7,-0.15);
\propag[boson] (-0.7,0.15) -- (0, 0.15) ;
\propag[boson] (0.7,0.15) -- (0, 0.15) ;
\filldraw [fill=white] (0,0) circle [radius=4mm] ;
\draw ($(0,0)+(90:0.4)$) to [out = -90, in = 10] ($(0,0)+(190:0.4)$) ;
\node (O1) at (-0.2,0.2) {$+$} ;
\node (O2) at (0.05,-0.05) {$-$} ;
\end{feynhand}
\end{tikzpicture}
}

\newcommand{\Auns}[1]{
\begin{tikzpicture}[baseline=-2]
\begin{feynhand}
\draw (0.7,-0.15) -- (-0.7,-0.15);
\propag[boson] (-0.7,0.15) -- (0, 0.15) ;
\propag[boson] (0.7,0.15) -- (0, 0.15) ;
\filldraw [fill=white] (0,0) circle [radius=4mm] ;
\node at (0,0) {$#1$} ;
\end{feynhand}
\end{tikzpicture}
}

\title{Anomalous Thresholds for the S-matrix of Unstable Particles }

\author[a]{Katsuki Aoki}
\author[b,c]{and Yu-tin Huang}

\affiliation[a]{Center for Gravitational Physics and Quantum Information, 
Yukawa Institute for Theoretical Physics, Kyoto University, 
606-8502, Kyoto, Japan}
\affiliation[b]{Department of Physics and Center for Theoretical Physics, National Taiwan University, Taipei
10617, Taiwan}
\affiliation[c]{Physics Division, National Center for Theoretical Sciences, Taipei 10617, Taiwan}

\abstract{In this work, we study the analytic properties of S-matrix for unstable particles, which is defined as the residues on the unphysical sheets where unstable poles reside. We demonstrate that anomalous thresholds associated with UV physics are unavoidable for unstable particles. This is in contrast to stable particles, where the anomalous thresholds are due to IR physics, set by the scale of the external kinematics. As a result, any dispersive representation for the amplitude will involve contributions from these thresholds that are not computable from the IR theory, and thus invalidate the general positivity bound. Indeed using toy models, we explicitly demonstrate that the four-derivative couplings for unstable particles can become negative, violating positivity bounds even for non-gravitational theories. Along the way, we show that contributions from anomalous thresholds in a given channel can be captured by the double discontinuity of that channel. }

\begin{document}
{\baselineskip0pt
\rightline{\baselineskip16pt\rm\vbox to-20pt{
           \hbox{YITP-23-167}
\vss}}%
}

\maketitle
\flushbottom

\section{Introduction}
The S-matrix bootstrap initiated in the middle of the last century began with the hope that the observable can be fully constrained by global symmetries, crossing, unitarity and analyticity.  However, as one can construct an infinite number of consistent quantum field theories (even demanding consistent coupling to gravity), each with its own S-matrix, it was quickly realized that the initial hope was somewhat misguided. On the other hand, while the S-matrix may not be unique, the region where the S-matrix is confined to can be viewed as the ``theory space'', in which all theories consistent with the previous principles must reside. This motivated the program of applying the bootstrap approach to the analysis of this space, which is conveniently parameterized by the Wilson coefficients of the low energy effective field theory (EFT) description.  Modern numerical methods and geometric understanding of the bootstrap equations have led to tremendous progress in delineating the boundary of this infinite dimensional space (see~\cite{Kruczenski:2022lot} and ~\cite{Baumgart:2022yty} for overview).

One of the most prominent examples is the positivity bound of the four-derivative coupling in any non-gravitational EFT~\cite{Adams:2006sv}. 
Such bounds have been applied to a wide range of phenomenological processes such as the chiral Lagrangian for pions ~\cite{Pham:1985cr, Manohar:2008tc}, WW scattering~\cite{Distler:2006if, Bellazzini:2014waa}, Higgs production~\cite{Low:2009di, Falkowski:2012vh}, and more recently standard model effective field theory (EFT)~\cite{Zhang:2020jyn, Remmen:2020vts, Li:2021lpe}. For derivation of such bounds from superluminal arguments see~\cite{CarrilloGonzalez:2022fwg}. In most of these cases, the operators in question often involve states that are not stable. However, their decay widths are usually under control in the weak coupling expansion and can be attributed to higher-order effects. On the other hand, it would be interesting to understand on general grounds, how the fact that the external states are unstable affects these positivity bounds.

At first glance, the problem appears to be ill-posed as by definition asymptotic states of the S-matrix must be stable. However since the presence of unstable particles can be detected from the presence of complex poles on the unphysical sheet of the usual S-matrix,  it was suggested long ago~\cite{Zwanziger:1963zza, Lvy1959OnTD, PhysRev.119.1121, PhysRev.123.692, Landshoff:1963nzy} that the S-matrix for unstable particles can be defined as the residue on these poles. Such a definition would allow one to infer the analytic properties of the unstable S-matrix by analytically continuing the unitarity equations for stable particles. Initial steps toward this direction were taken by one of the authors in~\cite{Aoki:2022qbf} (see also~\cite{Hannesdottir:2022bmo}), where the unitarity equation for unstable particles $2\rightarrow 2$ scattering was derived, taking a form very similar to that of stable-particle amplitudes.  However, as unitarity equations for stable particles are only applicable to physical kinematics, there exist regions of unstable kinematics that cannot be reached from physical kinematics. The aim of this paper is to close this gap.

For the purpose of positivity bounds, which is derived from the dispersive representation of the EFT coefficient and thus depends heavily on the non-analyticity of the amplitude near the forward limit, the pressing issue is when anomalous thresholds appear~\cite{Karplus:1958zz, Karplus:1959zz, Nambu:1958zze, RevModPhys.33.448}. Already for stable particles that are not the lightest state, anomalous thresholds appear on the physical sheet when $M>\sqrt{2}m$ (see \cite{Correia:2022dcu}). However since the presence of such singularities lies in the IR region defined by external kinematics, they can be computed via the EFT and safely subtracted, leaving terms that are constrained by the normal threshold in the dispersive representation.\footnote{Indeed one can even search for their presence in the collider~\cite{Boudjema:2008zn, Passarino:2018wix, Guo:2019twa}. We thank Sebastian Mizera for introducing these works.} Thus we would like to see if this continues to be the case for unstable particles.

We began by deriving the unitarity equations for the two stable two unstable particle S-matrix. This follows~\cite{Aoki:2022qbf} where one starts with the unitarity equations of stable particle scattering, and analytically continues the kinematics to complex values. Using $\pm$ to represent the decaying and growing mode of the unstable particle, we see that only for the conjugate setup $\mathcal{M}^{+-}_4$ do the unitarity equations agree with that of stable particles. For $\mathcal{M}^{++}_4$ we will have triangle singularities on the first sheet and no positivity can be established. 

We use explicit one-loop scalar integrals to verify the above result. For convenience, we present the scalar triangle and box integral in its dispersive representation to analyze its analytic continuation to complex kinematics. We find that anomalous thresholds do appear on the first sheet for both $\mathcal{M}^{+-}_4$  and $\mathcal{M}^{++}_4$! For $\mathcal{M}^{+-}_4$, we see that the triangle singularity enters the first sheet while Re$[t]>0$ which implies that it is reached by analytically continuing from the unphysical region, which explains why it was not observable in the previous analysis, i.e. the later utilizes unitarity equations of stable particles which only applies in the physical region. Furthermore, we find that the anomalous threshold can also occur when the internal state has a mass that is parametrically large compared to external kinematics! This is problematic since its source is associated with UV physics and this is not computable within the IR EFT, i.e. it is not subtractable. This implies that for unstable particles, \textit{on general grounds the positivity bounds should no longer be respected}. 

To better understand this conclusion we construct an explicit toy model where the system involves four distinct scalars $\pi, \phi, \chi_L, \chi_H$ with $\pi$ being the lightest state and the interaction given by:
\begin{align}
\mathcal{L}_{\rm int}= - \frac{g_{\phi}}{2} \phi \chi_L^2 - g_{\pi} \pi \chi_L \chi_H
\,.
\end{align}
We consider the four-point amplitude $\Amp_4(\pi(1) \phi(2) \phi(3) \pi(4))$. By tuning the mass of $\phi$ we can compare the situation between stable and unstable $\phi$. Assuming that the mass of $\chi_H$ is parametrically large compared to all other scales, we find that the coefficient of the four-derivative coupling $\tilde{B}_2$, which is normalized to be dimensionless, is given as 
$$\includegraphics[width=0.5\linewidth]{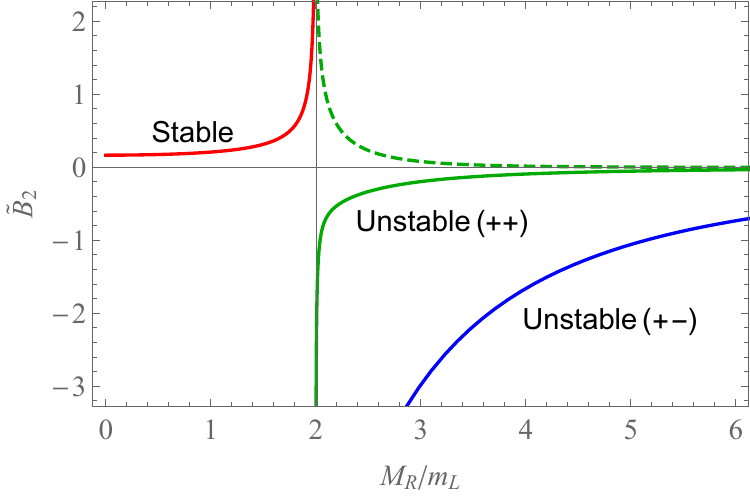}\,.$$
The horizontal axis is the mass ratio of the external mass to the internal mass, so the external state $\phi$ decays to $\chi_L$ above $M_R/m_L=2$.
We indeed find that the coefficient is negative for unstable kinematics even for the conjugate setup $\Amp^{+-}_4$! The choice $\Amp^{++}_4$ is worse as it is not even real (the solid and dashed curves are real and imaginary parts, respectively). To see that the negativity of $\Amp^{+-}_4$ is indeed due to anomalous thresholds, we consider the dispersive representation of the box integral for $\mathcal{M}^{+-}_4$, and use double discontinuity to isolate the triangle singularity which is the anomalous threshold:\footnote{Here the double discontinuity is with respect to the same variable, which is distinct from Mandelstam's double-discontinuity dispersive representation~\cite{PhysRev.115.1741}.}
\begin{align}
B_2^{+-}=\underbrace{\int_{m_{\rm th}^2}^{\infty} \frac{\D s}{2\pi i} \frac{2 \Disc_s \Amp^{+-} }{(s-M_R^2 -m^2)^3}}_{\mbox{$=\,\mathcal{I}^{+-}_n$: normal}}
+
\underbrace{\sum_n \int_{\mathcal{C}_n} \frac{\D s}{2\pi i} \frac{2 \Disc_s^2 \Amp^{+-} }{(s-M_R^2 -m^2)^3}}_{\mbox{$=\,\mathcal{I}^{+-}_a$: anomalous!}}
\,.
\end{align}
Each contribution after appropriate normalizations is given in the following: 
$$ \includegraphics[width=0.5\linewidth]{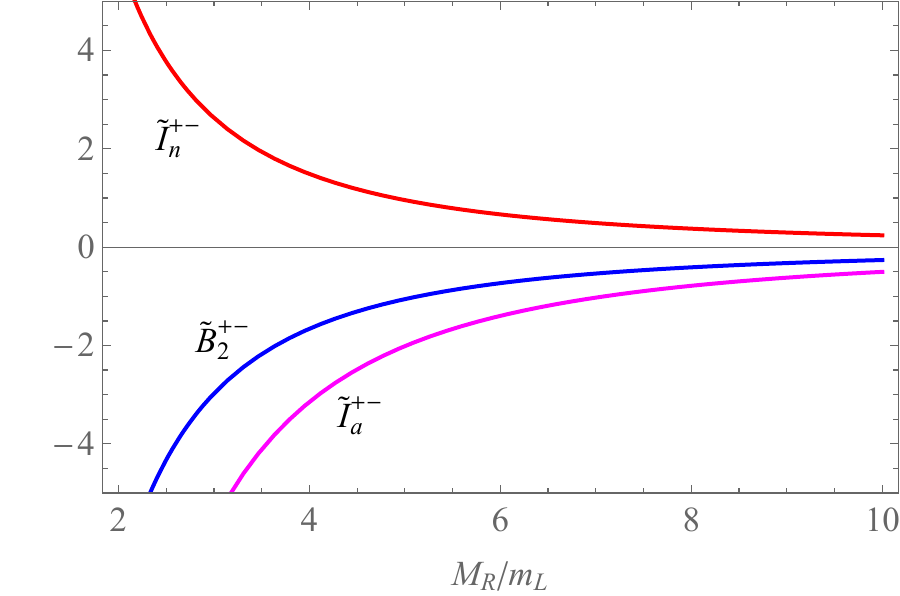}\,.$$
Indeed we find that the negative contribution comes solely from the anomalous threshold.

This paper is organized as follows: We start with a brief review of unstable particles and discuss properties coming from the factorizations of S-matrix in Sec.~\ref{sec:unstable}. In Sec.~\ref{sec:unitarity}, we use unitarity to see the analytic properties of unstable amplitudes. However, as is well-known, there are singularities that cannot be immediately seen from unitarity. Then, we study the analytic properties for unstable kinematics based on the explicit Feynmann diagrams in Sec.~\ref{sec:Feynman}. We elaborate on how anomalous thresholds appear on the first sheet when we change the mass of the external state and, importantly, we find anomalous thresholds at UV if the external state is unstable. In Sec.~\ref{sec:example}, we explicitly construct a model where the four-derivative coupling, which is defined by the low-energy expansion of the amplitude, becomes negative. In Sec.~\ref{sec:dispersion_relation}, we derive the dispersive representation and isolate the contribution from the UV anomalous thresholds by a double discontinuity formula, showing that the violation of the positivity bound comes from the UV anomalous thresholds. In Sec.~\ref{sec:Landau}, we dive into a generic one-loop diagram, discussing that the double discontinuity generically connects singularities at UV and IR. We conclude in Sec.~\ref{sec:conclusion}.

\section{Unstable particles and their S-matrix}
\label{sec:unstable}

We will begin with a brief discussion of the definition of unstable particles and their S-matrix (for recent review see~\cite{Hannesdottir:2022bmo}). Recall that given the 1 PI contribution to the two-point function $i\Sigma(p^2)$  the resumed propagator takes the form 
\begin{equation}
\frac{-i}{p^2 + m^2 -\Sigma(p^2)}=\frac{-i}{p^2 +M^2}
\end{equation} 
where $M^2$ is a shifted complex mass with 
\begin{equation}
{\rm Re}M^2=m^2-{\rm Re}\,\Sigma(p^2),\quad {\rm Im}\,M^2=-i   {\rm Im}\,\Sigma(p^2)\,.
\end{equation}
The imaginary part is identified with the decay rate $\Gamma= {\rm Im}\,\Sigma(p^2)/m$ leading to the Breit-Wigner shape for the distribution. Note that in general, we have $\Gamma>0$, i.e. the imaginary part of the mass is negative. Thus the on-shell condition for the unstable particle with the momentum $p^{\mu}$ should be understood as $p^2=-M^2$, such that in the rest frame the negative imaginary part implies that the ``wavefunction'' $e^{i p\cdot x} = e^{-iM t}$ decays in time so this is a decaying mode.

With the unstable particle defined, we now proceed to its S-matrix. Since resonances associated with unstable particles can be identified with poles on the unphysical sheet, see for example ~\cite{Zwanziger:1963zza} and~\cite{Aoki:2022qbf}, one can define the S-matrix with unstable external states as the residue of such poles. In the following, we will use amplitudes of identical real scalars and use their analytic properties to infer that of amplitudes for unstable particles.

\begin{figure}[h]
\centering
\includegraphics[width=0.7\linewidth]{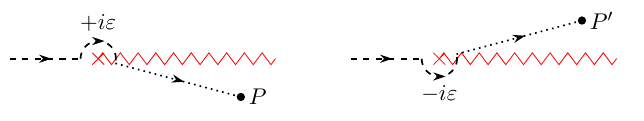}
\caption{Positions of complex poles where the zigzag lines denote branch cuts.}
\label{fig:complexpole}
\end{figure}

\begin{figure}[h]
\centering
\includegraphics[width=\linewidth]{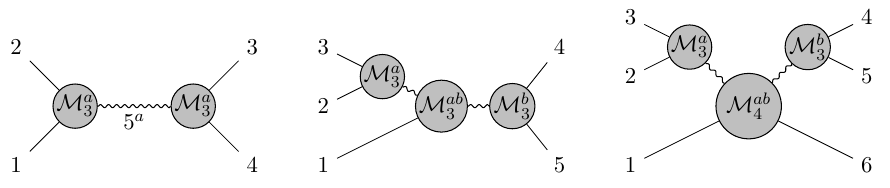}
\caption{Factorizations of amplitudes. The solid and wavy lines are stable and unstable particles, respectively.}
\label{fig:factorization}
\end{figure}

We begin with the scattering amplitude of stable particles satisfying real analyticity, $\Amp^*(s_A)=\Amp(s_A^*)$, where $s_A$ is the set of independent Lorentz-invariant variables. The 2-to-2 amplitude may have a resonance of the Breit-Wigner form which can be explained by a simple pole off the real axis, 
\begin{align}
\Amp_{2\to 2}(s,t) = \frac{\res{s=M^2} \Amp_{2\to 2} }{s-M^2} + \text{regular part}.
\end{align}
Here, $M^2$ is a complex number with a negative imaginary part and the position of the complex pole $P$ is shown by the left panel of Fig.~\ref{fig:complexpole}, where one analytically continues along the normal threshold and crosses over the branch cut to reach the unstable pole. The Mandelstam variables are defined by $s=-(p_1+p_2)^2, t= -(p_1+p_4)^2$ and $u=-(p_1+p_3)^2$, respectively. The resonance is interpreted as a production of a virtual unstable particle as illustrated by Fig.~\ref{fig:factorization} (left), or we can {\it define} unstable particles through the complex pole of the S-matrix. The position of the pole $M^2$ is identified as the complex mass squared of the unstable particle while the spin can be read off by expanding the residue in the center of mass frame on the Gegenbauer polynomials. 
The factorization property for the spin-0 particle implies
\begin{align}
\res{s=M^2} \Amp_{2\to 2} = -\Amp_3(1, 2, P^+) \Amp_3(3, 4, P^+)
\end{align}
where $\Amp_3(1, 2, P^+)$ is regarded as the on-shell three-point amplitude with two stable particles and one unstable particle. Here, the superscript $+$ represents the particle having a negative imaginary mass, i.e. a decaying mode.

The pole is reached by analytically continuing under the branch cut associated with physical thresholds. 
The real analyticity, $\Amp_{2\to 2}(s^*,t)=\Amp_{2\to 2}^*(s,t)$ for a fixed small $t$, requires another pole at the complex conjugate position $P'$ in the complex $s$-plane as shown by Fig.~\ref{fig:complexpole} (right). The on-shell condition reads $p^2=-(M^2)^*$ which has a positive imaginary part in its frequency, so the particle is a growing mode rather than decaying. This growing mode can be reached only if the amplitude is analytically continued from the anti-causal $-i\varepsilon$ direction. The residue at $P'$ is
\begin{align}
\res{s=(M^2)^*} \Amp_{2\to2} = -\Amp_3(1, 2, P^-) \Amp_3(3,4, P^-)
\end{align}
with
\begin{align}
\Amp_3(1, 2, P^-) = \Amp_3^* (1, 2, P^+)
\label{3point_conjugate}
\end{align}
Hereinafter, the superscripts $\pm$ denote whether the corresponding unstable particle is decaying ($+$) or growing ($-$). If no confusion arises, we omit the particle labels and write $\Amp_3^a = \Amp_3(1, 2, P^a)$ with $a=\pm$ and so on. 

So far we have defined the three-point amplitude with one unstable particle, where there is only one independent amplitude. The situation is different when amplitudes involve more than one unstable particle. Let us consider the $3 \to 2$ amplitude and extract the three-point amplitude with two unstable particles by utilizing the factorization\footnote{Throughout the paper, we use the notation $s_{ij\cdots}=-(p_i+p_j+\cdots)^2$ with $p_i$ being the momentum.}
\begin{align}
\Amp_{3\to 2} &\sim 
\begin{cases} 
\Amp_3^+ \frac{1}{s_{23}-M^2} \Amp_3^{++} \frac{1}{s_{45}-M^2} \Amp_3^+ &{\rm around}~s_{23}=M^2,~s_{45}=M^2
\,,
\\
\Amp_3^- \frac{1}{s_{23}-(M^2)^*} \Amp_3^{-+} \frac{1}{s_{45}-M^2} \Amp_3^+ &{\rm around}~s_{23}=(M^2)^*,~s_{45}=M^2
\,,
\\
\Amp_3^+ \frac{1}{s_{23}-M^2} \Amp_3^{+-} \frac{1}{s_{45}-(M^2)^*} \Amp_3^- &{\rm around}~s_{23}=M^2,~s_{45}=(M^2)^*
\,,
\\
\Amp_3^- \frac{1}{s_{23}-(M^2)^*} \Amp_3^{--} \frac{1}{s_{45}-(M^2)^*} \Amp_3^- &{\rm around}~s_{23}=(M^2)^*,~s_{45}=(M^2)^*
\,.
\end{cases}
\end{align}
See the middle panel of Fig.~\ref{fig:factorization}. Using the symmetry in the replacement $23 \leftrightarrow 45$ and the real analyticity $\Amp_{3\to 2}^*(s_{23}, s_{45}, \cdots) = \Amp_{3\to 2}(s_{23}^*, s_{45}^*, \cdots) $ where the ellipsis stands for the variables which are irrelevant to the residue, we find
\begin{align}
\Amp_3^{++}=(\Amp_3^{--})^*
\,, \quad
\Amp_3^{+-}=\Amp_3^{-+}=(\Amp_3^{+-})^*=(\Amp_3^{-+})^*
\,.
\label{amp3_relations}
\end{align}
There are now two independent three-point amplitudes, namely unstable particles are either the same or different. In particular, the symmetry $23 \leftrightarrow 45$ requires that the on-shell three-point amplitude with different choices $\Amp_3^{+-}$ is a real quantity while $\Amp_3^{++}$ is not necessarily real.

We proceed to discuss four-point amplitudes involving two external unstable particles, which is the main focus of this paper. We will define it through the six-point diagram of Fig.~\ref{fig:factorization}. We regard the six-point as a function of $s_{123}, s_{16}, s_{23}, s_{45}$ and variables that are irrelevant to the factorization. We keep the kinematics of the six-point amplitude fixed except for $s_{23}$ and $s_{45}$, which are continued above/below the cut for the decaying/growing mode. Note that in principle, due to momentum conservation the variables that are irrelevant to the factorization will also be deformed, which can lead to additional non-analyticity. This can result in the ambiguity in the definition of the unstable S-matrix. We will address this ambiguity in the next section. 

As in the previous case, the independent functions are $\Amp_4^{++}$ and $\Amp_4^{+-}$, and others can be obtained by complex conjugation, 
\begin{align}
 \Amp_4^{--}(s-i\varepsilon, t-i\varepsilon) &= \left(\Amp_4^{++}(s+i\varepsilon, t+i\varepsilon)\right)^*
\,, \nonumber\\
 \Amp_4^{+-}(s{+}i\varepsilon, t{+}i\varepsilon) = \Amp_4^{-+}(s{+}i\varepsilon, t{+}i \varepsilon)&=  \left(\Amp_4^{-+}(s{-}i\varepsilon, t{-}i\varepsilon)\right)^* = \left(\Amp_4^{+-}(s{-}i\varepsilon, t{-}i\varepsilon)\right)^*
\,,
\label{+-real_analyticity}
\end{align} 
where $\varepsilon \to +0$ is understood, and $s=s_{123}$ and $t=s_{16}$ are the Mandelstam variables of the embedded four-point amplitude. We have used the real analyticity and the symmetry in exchanges of external particles, $23 \leftrightarrow 45$ and $1\leftrightarrow 6$. 
Note that due to the multitude of complex invariants, the ``${\rm Im}$'' of the amplitude, defined as $\Amp-\Amp^*$, and the discontinuity in $s,t$, defined as $\Amp(s,t)-\Amp(s^*,t^*)$, may be different. Indeed we have, 
\begin{align}\label{eq: Dis}
\Disc \Amp^{++}_4(s, t) &=  \Amp^{++}_4(s+i \varepsilon, t+i \varepsilon) - \Amp^{++}_4(s-i \varepsilon, t-i \varepsilon ) \,, \nonumber\\
2i\im \Amp^{++}_4(s, t) &=  \Amp^{++}_4(s+i \varepsilon, t+i \varepsilon) - \Amp^{--}_4(s-i\varepsilon, t-i \varepsilon) 
\,, \nonumber\\
\Disc \Amp^{+-}_4(s, t) & =  2i \im \Amp^{+-}_4(s, t) =    \Amp^{+-}_4(s+i \varepsilon, t+ i\varepsilon) - \Amp^{+-}_4(s-i \varepsilon, t-i \varepsilon)
\,,
\end{align}
where $\Disc$ is the total discontinuity. The discontinuity and the imaginary part of $\Amp_4^{++}$ do not agree with each other, in general, while they agree in $\Amp_4^{+-}$.

It would be worth remarking that, in general, $\Amp_4^{++}$ and $\Amp_4^{--}$ are different functions although they are related by complex conjugation. In other words, they are not real analytic as seen from \eqref{eq: Dis}. The separation of the two can be understood by the presence of the external-mass singularity, which was studied thoroughly for the triangle diagram in Ref.~\cite{Hannesdottir:2022bmo}.\footnote{Indeed in Ref.~\cite{Hannesdottir:2022bmo}, it was shown that for unstable kinematics, the triangle diagram develops a cut that completely separates the upper and lower half-plane, with the function on the upper half related to the lower half via complex conjugation as can be seen from eq. (5.82) of~\cite{Hannesdottir:2022bmo}. } On the other hand, $\Amp_4^{+-}$ and $\Amp_4^{-+}$ are the same functions and real analytic. A path connecting upper-half and lower-half $s$-planes of $\Amp_4^{+-}$ is briefly discussed in~\cite{Aoki:2022qbf} under neglecting anomalous thresholds.

Amplitudes involving unstable external states also have their own threshold singularities. The discontinuity, defined in eq.~(\ref{eq: Dis}), admits a partial wave expansion where the expansion basis is dictated by Lorentz invariance to be Gegenbauer polynomials:
\begin{align}
\Disc \Amp_4^{ab}/2i= \sum_{\ell} \rho_{\ell}^{ab} G_{\ell}^{(D)} (\cos \theta)
\,.
\end{align}
For general complex kinematics the scattering angle $\theta$ is defined by
\begin{align}
\cos \theta :=1+ \frac{2(t-t_0)}{\sqrt{\lambda(s, s_1, s_2) \lambda(s, s_3 ,s_4)/s^2 } }
\,,
\label{def_cos}
\end{align}
with
\begin{align}
t_0=\frac{1}{2s}\left[ (s_1+s_2+s_3+s_4)s - s^2 +(s_1-s_2)(s_3-s_4) +  \sqrt{\lambda(s, s_1, s_2) \lambda(s, s_3 ,s_4)} \right]
\label{def_t0}
\end{align}
where $\lambda(x,y,z):=x^2+y^2+z^2-2xy-2yz-2zx$ is the K\"{a}ll\'{e}n function.
As the singularities correspond to physical threshold production, $\rho_{\ell}^{ab}$ should have a structure of the product of two three-point amplitudes with two lines being real masses and the third line being the complex mass. We now recall \eqref{3point_conjugate}; changing to the conjugate states yields the complex conjugate. We thus conclude that the discontinuity must be positively expandable on the Gegenbauer basis for the conjugate pair $\rho^{+-}_{\ell}>0$.
The exchanged state is not necessarily a one-particle state but may be a multi-particle state with angular momentum $\ell$. In this case, the states are continuously distributed and the singularity should be replaced with a brunch cut rather than a simple pole. 
This intuition, however, may fail if the partial wave expansion does not converge, i.e., $\Disc \Amp^{+-}$ has a singularity. Such a singularity is indeed a centrepiece of the present paper.

\section{Analytic properties of unstable scattering from (stable) unitarity}
\label{sec:unitarity}
In the previous section, we've seen that the discontinuity of the S-matrix for unstable particles associated with (unstable) threshold production is positively expandable on the Gegenbauer polynomials. However, as in stable particles, anomalous thresholds may appear and can be deduced from a repeated expansion of the unitarity equation (for example see chp.2 of ~\cite{Hannesdottir:2022bmo}). In this section, we use the physical unitarity equations together with the real analyticity to see what type of singularities can be detected. To simplify, we begin by considering a simple system of two scalars with $M> m$. For simplicity, we assume that $4m^2<M^2< 9m^2$ so that we only have two particle decays for the unstable particle.

We recall that the unstable-particle amplitudes are defined by residues of higher-point stable-particle amplitudes. Accordingly, analytic properties of the unstable-particle amplitudes can be obtained from unitarity constraints of higher-point stable-particle amplitudes. It is convenient to introduce the following diagrammatic notation~\cite{olive1964exploration,Eden:1966dnq}
\begin{align}
\begin{split}
\text{\scriptsize $n$}\Big\{
\begin{tikzpicture}[baseline=-2]
\draw (0.8,0.3) -- (-0.8,0.3);
\draw (0.8,0.1) -- (-0.8,0.1);
\draw [dotted] (0.8,-0.1) -- (-0.8,-0.1);
\draw (0.8,-0.3) -- (-0.8,-0.3);
\filldraw [fill=white] (0,0) circle [radius=5mm] ;
\node (O) at (0,0) {$\pm$} ;
\end{tikzpicture}
\Big\} \text{\scriptsize $n'$}
 &=-\Amp^{(\pm)}_{n \to n'}
 \,, \\
 \begin{tikzpicture}[baseline=-2]
\draw[line width=5pt] (-0.5, 0) -- (0.5,0); 
\end{tikzpicture}
&=
\sum_{a=2}^{\substack{\text{kinematically}\\ \text{allowed}}}
\begin{tikzpicture}[baseline=-2]
\draw (-0.5,0.3) -- (0.5,0.3) ;
\draw (-0.5,-0.3) -- (0.5,-0.3) ;
\draw (-0.5,0.1) -- (0.5, 0.1);
\draw[dotted] (-0.5, -0.1 ) -- (0.5,-0.1) ;
\end{tikzpicture}
\Big\} \text{\scriptsize $a$}
 \,, \\
 \text{each internal line} &= -2\pi i \theta(q^0) \delta(q^2+m^2)
 \,, \\
 \text{each loop} &= \frac{i}{(2\pi)^4} \int \D^4 k
 \,, 
 \\
 \text{$n$ lines joining two bubbles} &=\text{a symmetry factor } \frac{1}{n!}\,,
 \\
 \end{split}
 \label{rules}
\end{align}
where $\Amp^{(+)}_{n \to n'}$ is $n \to n'$ scattering amplitude for stable particles with mass $m$ where the $i \epsilon$ prescription is causal, and $\Amp^{(-)}_{n \to n'}$ is the complex conjugate. Each solid line denotes a single-particle state while the bold line is the sum of multi-particle states. For instance, the 2-to-2 unitarity equation is written as
\begin{align}
\tikzAmp{0.15,-0.15}{+} - \tikzAmp{0.15,-0.15}{-}
&=\tikzABA{0.15,-0.15}{+}{-}
\nn
&=
 \tikzAA{0.15, -0.15}{0.15,-0.15}{+}{-}
 +
 \tikzAA{0.15, -0.15}{0.2,0,-0.2}{+}{-}
+\cdots
\,.
\label{unitarity22}
\end{align}
The real analyticity states that the $(+)$ amplitude and the $(-)$ amplitude are the opposite boundary values of the same analytic function. Hence, LHS of \eqref{unitarity22} represents the discontinuity of the amplitude and the RHS tells us that it is due to the multi-particle intermediate states. Importantly, the unitarity equations are only applicable in the physical region. Thus in \eqref{unitarity22} for fixed Re $s$, only a finite set of intermediate states that are kinematically allowed to be on-shell contributes.  

The unitarity equation of the 3-to-3 scattering amplitude is obtained from the connected part of 
\begin{align}
\tikzLine{0.2,0,-0.2}
&=
\tikzABA{0.2, 0, -0.2}{S}{S^{\dagger}}
\nn
&=
\left(
\tikzLine{0.2,0,-0.2}
+
\sum
\tikzABsdis{+1}{0.2, 0}{-0.2}{0.1}{+}
+
\tikzAB{+1}{0.2,0,-0.2}{+}
\right)
\left(
\tikzLine{0.2,0,-0.2}
-
\sum
\tikzABsdis{-1}{0.2, 0}{-0.2}{0.1}{-}
-
\tikzAB{-1}{0.2,0,-0.2}{-}
\right)
\,,
\label{unitarity33}
\end{align}
where $S$ and $S^{\dagger}$ are the S-matrix elements which include both connected and disconnected diagrams. The summations are over possible choices of particles; for instance,
\begin{align}
\left(
\sum
\tikzABsdis{+1}{0.2, 0}{-0.2}{0.1}{+}
\right)
\left(
\sum
\tikzABsdis{-1}{0.2, 0}{-0.2}{0.1}{-}
\right)
&=
\underbrace{
\sum
\begin{tikzpicture}[baseline=-2]
\draw (-0.65,0.2) -- (-0.3,0.2);
\draw (-0.65,0) -- (-0.3,0);
\draw (0.65,0.2) -- (0.3,0.2);
\draw (0.65,0) -- (0.3,0);
\draw (-0.65,-0.2) -- (0.65,-0.2);
\draw[line width=5pt] (-0.3,0.1) -- (0.3,0.1);
\filldraw [fill=white] (-0.3,0.1) circle [radius=2mm] ;
\node at (-0.3,0.1) {$+$} ;
\filldraw [fill=white] (0.3,0.1) circle [radius=2mm] ;
\node at (0.3,0.1) {$-$} ;
\end{tikzpicture}
}_{\rm disconnected}
+
\underbrace{
\sum
\begin{tikzpicture}[baseline=-2]
\draw (-0.65,0.2) -- (0.65,0.2);
\draw (-0.65,0) -- (-0.3,0);
\draw (0.65,0) -- (0.3,0);
\draw (-0.65,-0.2) -- (0.65,-0.2);
\draw (-0.3,0.1) -- (0.3,-0.1);
\filldraw [fill=white] (-0.3,0.1) circle [radius=2mm] ;
\node at (-0.3,0.1) {$+$} ;
\filldraw [fill=white] (0.3,-0.1) circle [radius=2mm] ;
\node at (0.3,-0.1) {$-$} ;
\end{tikzpicture}
+
\sum
\begin{tikzpicture}[baseline=-2]
\draw (-0.65,0.2) -- (0.65,0.2);
\draw (-0.65,0) -- (-0.3,0);
\draw (0.65,0) -- (0.3,0);
\draw (-0.65,-0.2) -- (0.65,-0.2);
\draw[line width=5pt] (-0.3,0.1) -- (0.3,-0.1);
\filldraw [fill=white] (-0.3,0.1) circle [radius=2mm] ;
\node at (-0.3,0.1) {$+$} ;
\filldraw [fill=white] (0.3,-0.1) circle [radius=2mm] ;
\node at (0.3,-0.1) {$-$} ;
\end{tikzpicture}
}_{\rm connected}
\,,
\label{cross_connected}
\end{align}
where the sum sums over distinct assignments of the external legs to each blob, for example, there are in total 9 diagrams with a single internal line for the connected graph. The diagrams with multiple internal lines are kinematically forbidden when $4m^2<s_{\rm subenergy}<9m^2$ with $s_{\rm subenergy}=\{s_{12}, s_{23}, s_{13}, s_{45}, s_{56}, s_{46} \} $ for the label
\scalebox{0.6}{
\begin{tikzpicture}[baseline=-2]
\node (a) at (1,0.3) {4};
\node (b) at (1,0) {5};
\node (c) at (1,-0.3) {6};
\node (d) at (-1,0.3) {3};
\node (e) at (-1,0) {2};
\node (f) at (-1,-0.3) {1};
\draw (a) -- (d);
\draw (b) -- (e);
\draw (c) -- (f);
\filldraw [fill=white] (0,0) circle [radius=5mm] ;
\node (O) at (0,0) {} ;
\end{tikzpicture}
}. In the following, we will consider the scenario where $4m^2<s_{\rm subenergy}<9m^2$ since we are interested in the unstable particle that has 2-body decay only. The connected part of \eqref{unitarity33} then gives
\begin{align}
\tikzAmp{0.2,0,-0.2}{+} - \tikzAmp{0.2,0,-0.2}{-}
&=
\tikzABA{0.2,0,-0.2}{+}{-}
+\sum 
\tikzAAs{-1}{+}{-}
+\sum 
\tikzAAs{+1}{-}{+}
+ \sum
\tikzAAcross{+}{-}
\,.
\label{unitarity33c}
\end{align}
Note that the disconnected part of \eqref{unitarity33} independently holds thanks to the 2-to-2 unitarity.
Now an alternative expression of the 3-to-3 unitarity can be obtained by starting with the equation $SS^{\dagger}S=S$ rather than $SS^{\dagger}=1$:
\begin{align}
\adjincludegraphics[valign=c, width=0.8\linewidth]{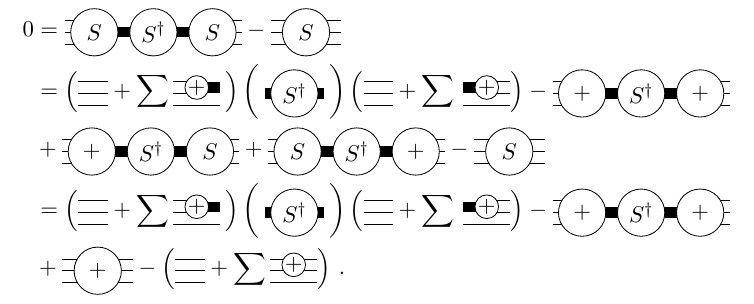}
\label{SSdaggerS}
\end{align}
Assuming $4m^2<s_{\rm subenergy}<9m^2$ and using the 2-to-2 unitarity, we can rewrite the first term in the last line as,
\begin{align}
\adjincludegraphics[valign=c, width=0.85\linewidth]{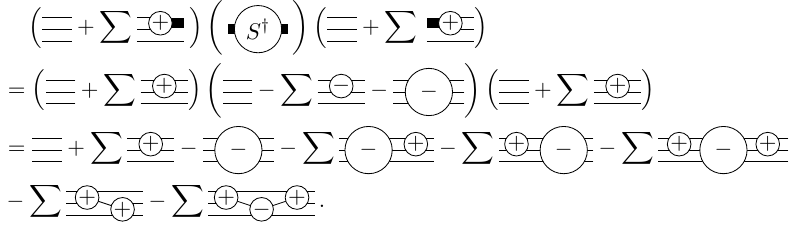}
\end{align}
Hence, the disconnected part of \eqref{SSdaggerS} is cancelled as it should be, whereas the connect part yields
\begin{align}
\adjincludegraphics[valign=c, width=0.8\linewidth]{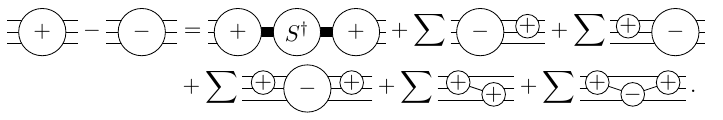}
\label{unitarity33plus}
\end{align}
The different forms of the 3-to-3 unitarity equations \eqref{unitarity33c} and \eqref{unitarity33plus} allow us to evaluate different discontinuities as we will see shortly.

The unitarity equations \eqref{unitarity33c} and \eqref{unitarity33plus} are composed of various terms and it is hard to see their implications. We adopt the proposal~\cite{olive1965unitarity} that each term of unitarity equations evaluates discontinuity across individual variables. It relies on the following postulated 2-particle discontinuity equations
\begin{align}
\begin{split}
\begin{tikzpicture}[baseline=-2]
\draw (0.6,0.2) -- (-0.6,0.2);
\draw (0.6,0) -- (-0.6,0);
\draw (0.6,-0.2) -- (-0.6,-0.2);
\filldraw [fill=white] (0,0) circle [radius=4mm] ;
\draw ($(0,0)+(90:0.4)$) to [out = -90, in = 170] ($(0,0)+(-10:0.4)$) ;
\node (O) at (0.2,0.2) {$+$} ;
\end{tikzpicture}
-
\begin{tikzpicture}[baseline=-2]
\draw (0.6,0.2) -- (-0.6,0.2);
\draw (0.6,0) -- (-0.6,0);
\draw (0.6,-0.2) -- (-0.6,-0.2);
\filldraw [fill=white] (0,0) circle [radius=4mm] ;
\draw ($(0,0)+(90:0.4)$) to [out = -90, in = 170] ($(0,0)+(-10:0.4)$) ;
\node (O) at (0.2,0.2) {$-$} ;
\end{tikzpicture}
&=
\begin{tikzpicture}[baseline=-2]
\draw (1,0.2) -- (-0.5,0.2);
\draw (1,0) -- (-0.5,0);
\draw (1,-0.2) -- (-0.5,-0.2);
\filldraw [fill=white] (0,0) circle [radius=4mm] ;
\draw ($(0,0)+(90:0.4)$) to [out = -90, in = 170] ($(0,0)+(-10:0.4)$) ;
\filldraw [fill=white] (0.7,0.1) circle [radius=2mm];
\node (O1) at (0.2,0.2) {$+$} ;
\node (O2) at (0.7,0.1) {$-$} ;
\end{tikzpicture}
=
\begin{tikzpicture}[baseline=-2]
\draw (1,0.2) -- (-0.5,0.2);
\draw (1,0) -- (-0.5,0);
\draw (1,-0.2) -- (-0.5,-0.2);
\filldraw [fill=white] (0,0) circle [radius=4mm] ;
\draw ($(0,0)+(90:0.4)$) to [out = -90, in = 170] ($(0,0)+(-10:0.4)$) ;
\filldraw [fill=white] (0.7,0.1) circle [radius=2mm];
\node (O1) at (0.2,0.2) {$-$} ;
\node (O2) at (0.7,0.1) {$+$} ;
\end{tikzpicture}
\,,  \\
\begin{tikzpicture}[baseline=-2]
\draw (0.6,0.2) -- (-0.6,0.2);
\draw (0.6,0) -- (-0.6,0);
\draw (0.6,-0.2) -- (-0.6,-0.2);
\filldraw [fill=white] (0,0) circle [radius=4mm] ;
\draw ($(0,0)+(90:0.4)$) to [out = -90, in = 10] ($(0,0)+(190:0.4)$) ;
\node (O) at (-0.2,0.2) {$+$} ;
\end{tikzpicture}
-
\begin{tikzpicture}[baseline=-2]
\draw (0.6,0.2) -- (-0.6,0.2);
\draw (0.6,0) -- (-0.6,0);
\draw (0.6,-0.2) -- (-0.6,-0.2);
\filldraw [fill=white] (0,0) circle [radius=4mm] ;
\draw ($(0,0)+(90:0.4)$) to [out = -90, in = 10] ($(0,0)+(190:0.4)$) ;
\node (O) at (-0.2,0.2) {$-$} ;
\end{tikzpicture}
&=
\begin{tikzpicture}[baseline=-2]
\draw (-1,0.2) -- (0.5,0.2);
\draw (-1,0) -- (0.5,0);
\draw (-1,-0.2) -- (0.5,-0.2);
\filldraw [fill=white] (0,0) circle [radius=4mm] ;
\draw ($(0,0)+(90:0.4)$) to [out = -90, in = 10] ($(0,0)+(190:0.4)$) ;
\filldraw [fill=white] (-0.7,0.1) circle [radius=2mm];
\node (O1) at (-0.2,0.2) {$-$} ;
\node (O2) at (-0.7,0.1) {$+$} ;
\end{tikzpicture}
=
\begin{tikzpicture}[baseline=-2]
\draw (-1,0.2) -- (0.5,0.2);
\draw (-1,0) -- (0.5,0);
\draw (-1,-0.2) -- (0.5,-0.2);
\filldraw [fill=white] (0,0) circle [radius=4mm] ;
\draw ($(0,0)+(90:0.4)$) to [out = -90, in = 10] ($(0,0)+(190:0.4)$) ;
\filldraw [fill=white] (-0.7,0.1) circle [radius=2mm];
\node (O1) at (-0.2,0.2) {$+$} ;
\node (O2) at (-0.7,0.1) {$-$} ;
\end{tikzpicture}
\,.
\end{split}
\label{discsub}
\end{align}
Here, the labels $(\pm)$ only refer to the ways of the approach of the specified subenergy variable, namely $s_{23}=-(p_2+p_3)^2$ and $s_{45}=-(p_4+p_5)^2$. All other variables are held at fixed real values. 
Applying the 2-particle discontinuity equations twice and using the 2-to-2 unitarity, we find
\begin{align}
\begin{tikzpicture}[baseline=-2]
\draw (0.7, 0.2) -- (-0.7, 0.2);
\draw (0.7, 0) -- (-0.7, 0);
\draw (0.7,-0.2) -- (-0.7,-0.2);
\filldraw [fill=white] (0, 0.0) circle [radius=4mm] ;
\draw ($(0,0)+(85:0.4)$) to [out = -90, in = 180] ($(0,0)+(350:0.4)$) ;
\draw ($(0,0)+(95:0.4)$) to [out = -90, in = 0] ($(0,0)+(190:0.4)$) ;
\node at (0.24,0.17) {$+$} ;
\node at (-0.24,0.17) {$+$} ;
\node at (0,-0.1) {$-$};
\end{tikzpicture}
-
\Athree{-}
=
\tikzAAs{-1}{-}{+} + \tikzAAs{+1}{-}{+} + \tikzAsAAs{+}{-}{+}
\,.
\end{align}
Let us reorganise the unitarity equations \eqref{unitarity33c} and \eqref{unitarity33plus} as 
\begin{align}
\tikzAsub{+}{+}{-} - \tikzAsub{+}{-}{-}
&=
\tikzABA{0.2,0,-0.2}{+}{-}
+ \sum \tikzAAcross{+}{-}
\,,
\label{unitarity33c2}
\\
\tikzAsub{+}{+}{+} - \tikzAsub{+}{-}{+}
&=
\tikzABABA{+}{S^{\dagger}}{+} + \sum \tikzAAcross{+}{+} + \sum \tikzTriangle{+}{-}{+}
\,,
\label{unitarity33plus2}
\end{align}
where on the LHS, the left/right entries refer to the position of the upper left/right two-particle subenergy variables while the centre entry refers to the remaining  variables~\cite{Eden:1966dnq}. More precisely,   
\begin{align}
\adjincludegraphics[valign=c, width=0.85\linewidth]{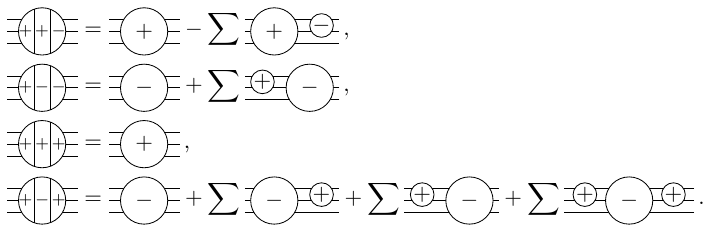}
\end{align}
For example in the first line, while $s_{23}$ is above the real axes, $s_{45}$ has been brought below the real axes by subtracting the two-particle discontinuity. Now, importantly, the left and right subenergies ($s_{23}$ and $s_{45}$) are aligned in \eqref{unitarity33c2} and \eqref{unitarity33plus2}. Therefore, \eqref{unitarity33c2} and \eqref{unitarity33plus2} are understood as the discontinuities for fixed subenergy variables, which can be used to define the unitarity equation for unstable particles!

As mentioned in the previous section, the continuation in $s_{23}$ and $s_{45}$ will inevitably deform other Mandelstam variables which could encounter additional thresholds. This leads to an ambiguity, i.e. the unstable S-matrix which requires us to move off the physical sheet is ambiguous as it depends on ``how'' the analytic continuation is done. This is not surprising as the amplitude is multi-sheeted and a given unstable pole can appear on different sheets. As we would like to infer the property of unstable particle S-matrix from unitarity, we assume the existence of a certain region of kinematics such that a complex pole is the singularity closest to the real axis. This particularly means that the loop integral of the unitarity equations does not require contour deformation during the analytic continuation. We define the unstable-particle amplitude $\Amp_4^{ab}$ in such a region and discuss its unitarity equation. We will come back to the issue of other singularities at the end of this section.

We analytically continue \eqref{unitarity33c2} and \eqref{unitarity33plus2} in the variables $s_{23}$ and $s_{45}$ for fixed $s$ and $t$ where the variables $(s,t,u)$ are the Mandelstam variables of the embedded 2-to-2 diagram. They are related to the variables of the 3-to-3 diagram via $s=s_{123}=s_{456}, t=s_{16}=s_{2345}$ and $u=s_{145}=s_{236}$. Note that the second term of RHS of \eqref{unitarity33c2} and \eqref{unitarity33plus2} includes the $u$-channel single-particle exchange diagram [see \eqref{cross_connected}]. However, the diagram is proportional to $\delta(m^2-u)$ so we can ignore the $u$-channel diagram as long as $u=2m^2+s_{23}+s_{45}-(s+t)\neq m^2$. Then, the analytic continuations of \eqref{unitarity33c2} and \eqref{unitarity33plus2} give, after divided by common factors,\footnote{The unitarity equation of $\Amp_4^{++}$ was briefly discussed in~\cite{Aoki:2022qbf} in a slightly different way. The expression in~\cite{Aoki:2022qbf} is not symmetrical in the in/out states while \eqref{unitarity22unsplus} has such a symmetry, but two expressions are equivalent under unitarity.}
\begin{align}
 \AmixL - \AmixR
&=
\begin{tikzpicture}[baseline=-2]
\begin{feynhand}
\draw (-1.2,-0.2) -- (-0.5,-0.2);
\draw (1.2,-0.2) -- (0.5,-0.2);
\draw[line width=5pt] (-0.5, 0) -- (0.5,0);
\propag [boson] (-1.2,0.1) -- (-0.5, 0.1) ;
\propag [boson] (1.2,0.1) -- (0.5, 0.1) ;
\filldraw [fill=white] (-0.5,0) circle [radius=4mm] ;
\node (O1) at (-0.5,0) {$+$} ;
\filldraw [fill=white] (0.5,0) circle [radius=4mm] ;
\node (O2) at (0.5,0) {$-$} ;
\end{feynhand}
\end{tikzpicture}
\label{unitarity22uns}
\\
\Auns{+}
-
\begin{tikzpicture}[baseline=-2]
\begin{feynhand}
\propag[boson] (-0.7,0.15) -- (0, 0.15) ;
\propag[boson] (0.7,0.15) -- (0, 0.15) ;
\draw (0.7,-0.15) -- (-0.7,-0.15);
\filldraw [fill=white] (0, 0.0) circle [radius=4mm] ;
\draw ($(0,0)+(85:0.4)$) to [out = -90, in = 180] ($(0,0)+(0:0.4)$) ;
\draw ($(0,0)+(95:0.4)$) to [out = -90, in = 0] ($(0,0)+(180:0.4)$) ;
\node at (0.24,0.2) {$+$} ;
\node at (-0.24,0.2) {$+$} ;
\node at (0, -0.05) {$-$};
\end{feynhand}
\end{tikzpicture}
&=
\begin{tikzpicture}[baseline=-2]
\begin{feynhand}
\propag[boson]  (-1.2,0.1) -- (-0.5,0.1);
\draw (-1.2,-0.2) -- (-0.5,-0.2);
\draw (2.2,-0.2) -- (1.5,-0.2);
\propag[boson]  (2.2,0.1) -- (1.5,0.1);
\draw[line width=5pt] (-0.5,0) -- (1.5,0);
\filldraw [fill=white] (-0.5,0) circle [radius=4mm] ;
\node (O1) at (-0.5,0) {$+$} ;
\filldraw [fill=white] (0.5,0) circle [radius=4mm] ;
\node (O2) at (0.5,0) {$S^{\dagger}$} ;
\filldraw [fill=white] (1.5,0) circle [radius=4mm] ;
\node at (1.5,0) {$+$} ;
\end{feynhand}
\end{tikzpicture}
+
\begin{tikzpicture}[baseline=-2]
\begin{feynhand}
\draw (1.1,-0.2) -- (-1.1, -0.2) ;
\draw (0.6,0.2) -- (-0.6,0.2);
\propag[boson] (1.1,0.1) -- (0.6,0.1);
\propag[boson] (-1.1,0.1) -- (-0.6,0.1);
\draw (-0.6,0.1) -- (0,-0.1);
\draw (0.6,0.1) -- (0,-0.1);
\filldraw [fill=white] (-0.6,0.1) circle [radius=2mm] ;
\node at (-0.6,0.1) {$+$} ;
\filldraw [fill=white] (0,-0.1) circle [radius=2mm] ;
\node at (0,-0.1) {$-$} ;
\filldraw [fill=white] (0.6,0.1) circle [radius=2mm] ;
\node at (0.6,0.1) {$+$} ;
\end{feynhand}
\end{tikzpicture}
\,,
\label{unitarity22unsplus}
\end{align}
with wavy lines being unstable particles and 
\begin{align}
\AmixL&=-\Amp_4^{+-}(s+i\varepsilon, t)
\,, \quad
\AmixR=-\Amp_4^{+-}(s-i\varepsilon, t)
\,. \\
\Auns{+}&=-\Amp_4^{++}(s+i\varepsilon, t+i\varepsilon)
\,, \quad
\begin{tikzpicture}[baseline=-2]
\begin{feynhand}
\propag[boson] (-0.7,0.15) -- (0, 0.15) ;
\propag[boson] (0.7,0.15) -- (0, 0.15) ;
\draw (0.7,-0.15) -- (-0.7,-0.15);
\filldraw [fill=white] (0, 0.0) circle [radius=4mm] ;
\draw ($(0,0)+(85:0.4)$) to [out = -90, in = 180] ($(0,0)+(0:0.4)$) ;
\draw ($(0,0)+(95:0.4)$) to [out = -90, in = 0] ($(0,0)+(180:0.4)$) ;
\node at (0.24,0.2) {$+$} ;
\node at (-0.24,0.2) {$+$} ;
\node at (0, -0.05) {$-$};
\end{feynhand}
\end{tikzpicture}
=-\Amp_4^{++}(s-i\varepsilon, t-i\varepsilon)
\,.
\end{align}
As for $\Amp_4^{+-}$, we have not added $i\varepsilon$ to $t$ because \eqref{unitarity33c2} [or \eqref{unitarity22uns}] suggests no $t$-channel type singularity in the region of our interest.

Let us compare the two discontinuities in \eqref{unitarity22uns} and \eqref{unitarity22unsplus}. One important difference is the absence/existence of the $t$-channel triangle diagram. To understand it, we should recall that the unitarity equations are applied in the physical region which requires $t=s_{16}<0$ in the case of 3-to-3 amplitude. This suggests that the absence/existence of the $t$-channel triangle singularity stems from where the unstable kinematics is continued from. In the next section, we will study the analytic properties of explicit Feynman integrals to verify this analysis. 


The analysis in this section should be considered as proof of existence, i.e. contributions to the discontinuities that are implied from the unitarity equations. It does not imply \textit{the absence} of particular singularities. Indeed, when continuing the external kinematics into unstable regions, many new singularities might arise along the way. For example, some singularities ``hidden'' in a bubble on the RHS of unitarity equations may generate a new singularity of LHS through the analytic continuation. For instance, as we will see in the next section, the perturbative analysis of $\Amp_4^{+-}$ suggests there exist $s$-channel (and $u$-channel) triangle singularities at complex positions (i.e.~away from the physical region) which cannot be immediately seen from \eqref{unitarity22uns} and \eqref{unitarity22unsplus}.

\begin{figure}[t]
\centering
\includegraphics[width=1\linewidth]{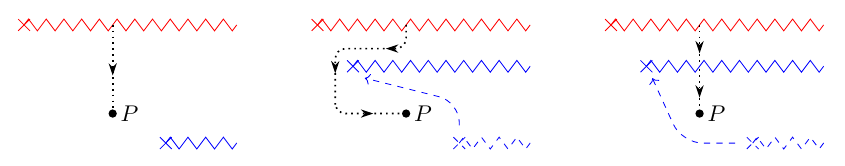}
\caption{Paths to reach the complex pole in the presence of additional singularities denoted by the blue zigzag lines. The blue arrows represent how the branch point moved from its original position (dashed zigzag lines) by changing the kinematics.}
\label{fig:complexpole2}
\end{figure}

Let us be a little more concrete. So far, we are considering a continuation such that other singularities are away from the path (the left panel of Fig.~\ref{fig:complexpole2}). It is easy to imagine other continuations where additional singularities move closer to the real axes (the middle and right panels of Fig.~\ref{fig:complexpole2}). Depending on the continuation, the branch point of the new singularity can approach the real axis by moving above the complex pole (the middle) or below it (the right). For the former, one simply deforms the original continuation path, while for the latter one inevitably crosses the branch cut of the new singularity. Thus the ambiguity of the definition of unstable pole $P$ reflects the multi-valuedness of the amplitude with respect to multiple invariants. The middle and right figures actually talk about the same complex pole but for different sheets of kinematic invariants. Due to the extra singularity, one cannot directly infer properties of the residue for the complex pole from continuing from unitarity equations alone. We need a knowledge of positions of such extra singularities.

\section{Explicit Feynman diagram analysis}
\label{sec:Feynman}

\begin{figure}[h]
\centering
\includegraphics[width=0.7\linewidth]{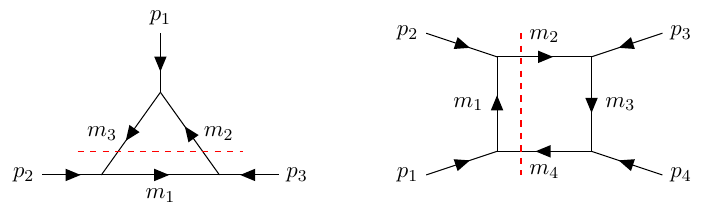}
\caption{The triangle diagram and the box diagram. The dashed lines denote the cuts relevant to discontinuities.}
\label{fig:tri_box}
\end{figure}

Let us discuss the analytic structure of the 2-to-2 unstable-particle amplitudes at the one-loop level. The relevant one-loop diagrams for the 2-to-2 amplitudes are the bubble diagram (2-point diagram), the triangle diagram (3-point diagram), and the box diagram (4-point diagram). Since the bubble diagram has no anomalous threshold, we shall focus on the triangle and box diagrams with the internal mass and the momenta shown by Fig.~\ref{fig:tri_box}.

Using the Feynman parameters, we have
\begin{align}
{\cal I}_{\rm tri}(s_i, m_i^2) &=\int^1_0 \left[ \prod_{i=1}^3 \D \alpha_i \right] \delta(1-\sum_{i=1}^3 \alpha_i ) \frac{1}{D_{\rm tri}}
\,, 
\label{Itri}\\
{\cal I}_{\rm box}(s_{12}, s_{23}, s_i, m_i^2) &= \int^1_0 \left[ \prod_{i=1}^4 \D \alpha_i \right] \delta(1-\sum_{i=1}^4 \alpha_i ) \frac{1}{D_{\rm box}^2}
\,,
\label{Ibox}
\end{align}
where
\begin{align}
s_i&:= -p_i^2\,, \quad
s_{ij}:=-(p_i+p_j)^2\,
\end{align}
and
\begin{align}
D_{\rm tri}&=\alpha_2 \alpha_3 s_1 + \alpha_3 \alpha_1 s_2 + \alpha_1 \alpha_2 s_3 -\left( \sum_{i=1}^3 \alpha_i m_i^2 \right) (\alpha_1+\alpha_2+\alpha_3)
\,, \\
D_{\rm box}&=\alpha_2 \alpha_4 s_{12}+\alpha_1 \alpha_3 s_{23}+\alpha_4 \alpha_1 s_1 + \alpha_1 \alpha_2 s_2 +\alpha_2 \alpha_3 s_3 +\alpha_3 \alpha_4 s_4 
\nn
&-\left( \sum_{i=1}^4 \alpha_i m_i^2 \right) (\alpha_1+\alpha_2+\alpha_3+\alpha_4)\,.
\end{align}
Here, we focus on four dimensions for simplicity. For unstable particles, $p_i$ can be interpreted as the internal momentum of a higher-point amplitude, but analytic continued to complex values.  

In principle, the locations of the singularities can be analysed by the Landau equations. However, this will not be suitable for our purpose since the Landau equations only give necessary conditions, and furthermore,  we will be considering complex external kinematics. Instead, we use the dispersive representation starting with a region where analyticity is known and then analytically continue. Since we will be interested in the two unstable two stable particle scattering, we will analytically continue two of the external kinematics, one after the other, to complex values.

\subsection{Anomalous thresholds for stable particles}
\label{sec:anomalous1loop}

It is well known that anomalous thresholds can appear on the physical sheet for stable particles that are not the lightest state in the spectrum. Indeed this can be seen from the analysis of Landau equations. In this section, we will instead proceed with the dispersive representation of Feynman integrals, demonstrating the presence of anomalous thresholds. The advantage is that we can analytically continue the external kinematics in the representation, and detect when the triangle singularity enters the first sheet.

Let's begin with the dispersive representation for the triangle and box integrals, which do not require subtraction terms in four-dimensions. Fixing all the Mandelstam variables except for $s_1$ for the triangle and $s_{12}$ for the box diagram, we will consider the physical scattering process in the $s_1$-channel and the $s_{12}$-channel respectively. For instance, in the case of the triangle diagram, as long as $s_2$ and $s_3$ are small enough (for example, one can consider the case $s_2\simeq s_3 \simeq m_1^2 \simeq m_2^2 \simeq m_3^2$), the only singularities of $\mathcal{I}_{\rm tri}$ on the first sheet is the normal threshold starting at $s_1=(m_2+m_3)^2$. Similarly, the box integral only has the normal threshold if $s_{23}$ and $s_i$ are fixed in sufficiently small values. Then, the integrals may be written as
\begin{align}
\mathcal{I}_{\rm tri}&=\frac{1}{2\pi i} \int_{(m_2+m_3)^2}^{\infty} \D s_{1}' \frac{ \Disc_{s_{1}} \mathcal{I}_{\rm tri}(s'_{1})}{s'_{1}-s_{1} }
=
\int_{(m_2+m_3)^2}^{\infty} \D s_{1}' \frac{ \rho_{\rm tri}(s'_{1})}{s'_{1}-s_{1} }
\,, 
\label{disp_tri} \\
\mathcal{I}_{\rm box}&=\frac{1}{2\pi i} \int_{(m_2+m_4)^2}^{\infty} \D s_{12}' \frac{ \Disc_{s_{12}} \mathcal{I}_{\rm box}(s'_{12})}{s'_{12}-s_{12} }
=
\int_{(m_2+m_4)^2}^{\infty} \D s_{12}' \frac{ \rho_{\rm box}(s'_{12})}{s'_{12}-s_{12} }
\,,
\label{disp_box}
\end{align}
with
\begin{align}
\rho_{\rm tri}&=\frac{1}{\lambda^{1/2}(s_1,s_2,s_3)} 
\ln\left[ \frac{ \sqrt{ 2s_1 S_{\rm tri} + \lambda(s_1, m_2^2, m_3^2) \lambda(s_1, s_2, s_3)  }- \sqrt{\lambda(s_1, m_2^2, m_3^2) \lambda(s_1,s_2,s_3)} }
{ \sqrt{2s_1 S_{\rm tri} + \lambda(s_1, m_2^2, m_3^2) \lambda(s_1, s_2, s_3) } + \sqrt{\lambda(s_1, m_2^2, m_3^2) \lambda(s_1,s_2,s_3)}   }
\right]
\label{rho_tri}
\\
\rho_{\rm box}&=\frac{1}{S_{\rm box}^{1/2}} 
\ln\left[ \frac{ \sqrt{S_{\rm tri}^L S_{\rm tri}^R+\lambda(s_{12}, m_2^2,m_4^2) S_{\rm box} } + \sqrt{\lambda(s_{12}, m_2^2, m_4^2) S_{\rm box}} }
{ \sqrt{S_{\rm tri}^L S_{\rm tri}^R+\lambda(s_{12}, m_2^2,m_4^2) S_{\rm box} } - \sqrt{\lambda(s_{12}, m_2^2, m_4^2) S_{\rm box}}  }
\right]
. \label{rho_box}
\end{align}
Here, we have introduced $\lambda(x,y,z):=x^2 + y^2 + z^2 - 2xy - 2yz - 2zx$, 
\begin{align}
S_{\rm tri}& := {\rm det}\frac{\partial D_{\rm tri}}{\partial \alpha_i \partial \alpha_j} \quad (i,j=1,2,3)\,, \\
S_{\rm box}& := {\rm det}\frac{\partial D_{\rm box}}{\partial \alpha_i \partial \alpha_j} \quad (i,j=1,2,3,4)\,,
\end{align}
and
\begin{align}
S_{\rm tri}^L &:= {\rm det}\frac{\partial D_{\rm box}}{\partial \alpha_i \partial \alpha_j} \quad (i,j=1,2,4)\,, \\
S_{\rm tri}^R &:= {\rm det}\frac{\partial D_{\rm box}}{\partial \alpha_i \partial \alpha_j} \quad (i,j=2,3,4)\,.
\end{align}
which should satisfy
\begin{align}
\lambda(s_1, m_2^2,m_3^2)>0\,, \quad S_{\rm tri}&>0\,, \quad  \lambda(s_1,s_2,s_3)>0\,, \\
\lambda(s_{12}, m_2^2,m_4^2)>0\,, \quad S_{\rm box}&>0\,, \quad S_{\rm tri}^L>0\,, \quad S_{\rm tri}^R>0\,, 
\end{align}
along the dispersive integral, i.e. $(m_2+m_3)^2<s_1<\infty$ and $(m_2+m_4)^2<s_{12}<\infty$. Hence, $\rho_{\rm tri}$ and $\rho_{\rm box}$ are real and finite along the contour in eq.(\ref{disp_tri}) and eq.(\ref{disp_box})

\begin{figure}[h]
\centering
\includegraphics[width=0.6\linewidth]{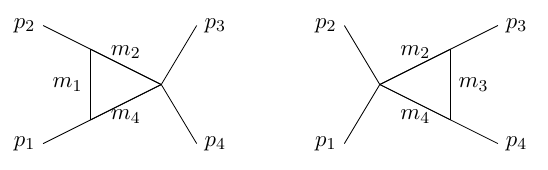}
\caption{The relevant reduced diagrams for the $s_{12}$ cut of the box diagram.}
\label{fig:reduced}
\end{figure}

Let us now analytically continue $\mathcal{I}_{\rm tri}$ or $\mathcal{I}_{\rm boxi}$ from the original domain. A singularity of the integral $\mathcal{I}_A$ arises when the integration contour is pinched between a singularity of $\rho_A~(A={\rm tri}, {\rm box})$ and the singularity due to the denominator of \eqref{disp_tri} or \eqref{disp_box}.\footnote{This is the singularity on the $s$-plane. Singularities of the integral $\mathcal{I}_A$ in other variables are found when singularities of $\rho_A$ pinch the integration contour or a singularity of $\rho_A$ touches the endpoint of the integral. They will be studied in Sec.~\ref{sec:Landau}.} The conditions $S_{\rm tri}=0$ and $S_{\rm box}=0$ are the conditions for the leading singularities of the triangle and box diagram while $S_{\rm tri}^L=0$ and $S_{\rm tri}^R=0$ are the lower-order singularities associated with the reduced diagrams shown in Fig.~\ref{fig:reduced}.

Notice that since the $\rho_A$s are multi-valued functions, the singularities of $\rho_A$ are only relevant if they appear on the first sheet. For example, $\rho_{\rm tri}$ may be singular at $\lambda(s_1,s_2,s_3)=0$ but it is not the case when the logarithm takes the principal value. The triangle singularities $S_{\rm tri}=0, S_{\rm tri}^L=0, S_{\rm tri}^R=0$ are always singularities of $\rho_A$ at which the argument of the logarithm is either 0 or complex infinity. For simplicity, we will consider $S_{\rm tri}=0$ in the following.

The equation $S_{\rm tri}=0$ is quadratic in $s_1$ and there are two independent roots:
\begin{align}
s_1 = s_{\pm} := s_2+s_3 + \frac{-(s_3+m_1^2-m_2^2)(s_2+m_1^2-m_3^2) \pm \sqrt{ \lambda(s_2, m_1^2, m_3^2) \lambda(s_3, m_1^2, m_2^2)} }{2m_1^2}
\,.
\label{s1roots}
\end{align}
The question is now when does this singularity pinch the contour on the first sheet. Since the contour is along  $(m_2{+}m_3)^2<s_1<\infty$, in order for the contour to be pinched, the triangle singularity  $s_1=s_{\pm}$ must first come to the vicinity of the branch point $s_1=(m_2{+}m_3)^2$. As $ s_{\pm} $ is a function of $s_2, s_3$ and internal masses, the detail of when a pinch occurs depends heavily on the regions of these parameters. 

Let us begin with the degenerate case where we set both external kinematics to be the same $s_2=s_3=M^2$ and the internal mass to be identical $m_i=m$. Then the solution to $S_{\rm tri}=0$ becomes $s_1=(0,\frac{M^2(4m^2-M^2)}{m^2})$. As we increase $M^2$ the second triangle singularity moves closer to the branch point $s_1=4m^2$. At $M=\sqrt{2}m$ it reaches the branch point and can move onto the physical sheet, leading to the famous anomalous threshold.

\begin{figure}[h]
\centering
\includegraphics[width=\linewidth]{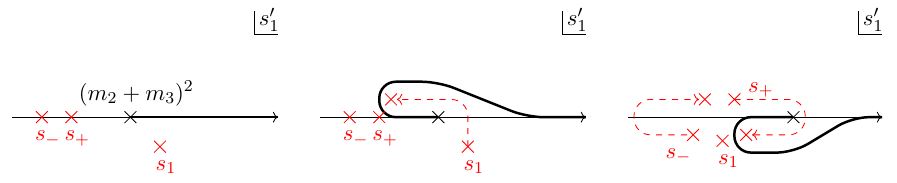}
\caption{Generations of triangle singularities. The $\times$ are the singularities of the integrand and the thick curves are the integration contour. Left: $s_{\pm}$ and $s_1$ do not pinch the contour, showing no triangle singularity on the first sheet. Middle: As $s_1$ moves through the branch cut associated with the normal threshold, the contour is distorted and a pinch occurs, that is, $\mathcal{I}_{\rm tri}$ has a triangle singularity at $s_1=s_{\pm}$ on the second sheet. Right: $s_+$ moves and go round $(m_2+m_3)^2$ as $s_2$ and/or $s_3$ increase. Then, a singularity is found at $s_1=s_+$ on the first sheet while $s_1=s_-$ remains regular. This is the usual anomalous threshold for stable particles. Note that the figures are drawn to understand qualitative behaviours and not drawn by using \eqref{s1roots}. }
\label{fig:generation}
\end{figure}

Similar results can be obtained for general mass distribution. We begin with $s_2=(m_1-m_3)^2 + \epsilon$, $s_3=(m_1-m_2)^2 + \epsilon$ with a small positive $\epsilon$ at which the roots are real and $s_{\pm}\simeq (m_2-m_3)^2$. The singularities $s_1'  = s_{\pm} \simeq (m_2-m_3)^2$ are outside of the integration contour $(m_2+m_3)^2<s_1'<\infty$ as displayed in the left of Fig.~\ref{fig:generation}. For this to become a singularity of the integral, we need to move $s_1$ across the branch cut in order to pinch the integration contour as shown in the center of Fig.~\ref{fig:generation}, and thus the singularity lives on the second sheet. As $s_2$ and/or $s_3$ increase, the singularities $s_{\pm}$ move along the real $s_1'$ axis and $s_+$ will touch the end point of the integration $s_1'=(m_2+m_3)^2$ when $m_2 s_2 +m_3 s_3 - (m_1^2+m_2 m_3)(m_2+m_3) = 0$. We add a small imaginary part to bypass the endpoint singularity,
\begin{align}\label{Conditions}
m_2 s_2 +m_3 s_3 >  (m_1^2+m_2 m_3)(m_2+m_3) \,.
\end{align}
In such case, the integration contour will be deformed in such a way that it can be pinched at  $s_1=s_+$ without $s_1$ passing through the branch cut as illustrated on the right of Fig.~\ref{fig:generation}. Thus the triangle singularity of $\mathcal{I}_{\rm tri}(s_1)$ arises on the first sheet at $s_1=s_+$. Importantly, note that if any of the internal mass is parametrically large compared to the external kinematics, the inequality cannot be satisfied. Thus the anomalous threshold here is strictly IR, and can be reliably computed and subtracted.

\subsection{Anomalous thresholds for unstable particles: IR}

With the anomalous threshold on the first sheet, we further increase $s_2$ and/or $s_3$ and eventually, we will obtain the anomalous threshold for unstable configurations. We will follow the migration of the roots to track the kinematic region where the threshold resides for different unstable configurations. This will help us elucidate how from the unitarity equations, the triangle singularity only appeared in one particular unstable kinematics ($+,+$) and not the other ($+,-$). The two roots $s_{\pm}$ will eventually become degenerate at either $s_2=(m_1+m_3)^2$ or $s_3=(m_1+m_2)^2$, i.e., at the decay threshold. We can go round the threshold by adding a small imaginary part (cf. Fig~\ref{fig:complexpole}) to avoid the pinch between $s_{\pm}$ and then reach the region where $s_2$ and/or $s_3$ are unstable.

Keeping $s_3$ stable, i.e., $\lambda(s_3, m_1^2, m_3^2)=(s_3{-}(m_1{-}m_2)^2)(s_3{-}(m_1{+}m_2)^2)<0$, we study the neighbourhood of the decaying threshold, $s_2=(m_1+m_3)^2+\delta s_2$, in which the K\"{a}ll\'{e}n function takes
\begin{align}
\lambda(s_2,m_1^2, m_3^2)=4m_1m_3 \delta s_2 + \mathcal{O}(\delta s_2^2)
\,.
\end{align}
To bypass the decaying threshold, $s_2$ needs to be rotated more than $\pi$ but less than $3\pi/2$ (see Fig.~\ref{fig:complexpole}) and thus $s_{\pm}$ after the continuation are given by
\begin{align}
 s_{\pm} \to s_2+s_3 + \frac{-(s_3+m_1^2-m_2^2)(s_2+m_1^2-m_3^2) \mp \sqrt{ \lambda(s_2, m_1^2, m_3^2) \lambda(s_3, m_1^2, m_2^2)} }{2m_1^2}
 \,.
 \label{spmafter1}
\end{align}
where the square root is understood as the principal square root. The root $s_{+}$ is complex even if $s_2$ is almost real, namely the narrow width approximation for $s_2$. For $s_2=P$ (decaying mode), the triangle singularity $s_1=s_+$ appears in the lower-half plane and it exists in the upper-half plane for $s_2=P'$ (growing mode).

The expression \eqref{spmafter1} remains valid even after $s_3$ also goes beyond the decaying threshold if both $s_2$ and $s_3$ bypass the thresholds from the same complex direction, i.e., both $s_2$ and $s_3$ follow the same path of Fig.~\ref{fig:complexpole}. On the other hand, if $s_2$ and $s_3$ bypass the thresholds from opposite directions (one is the $+i\varepsilon$ path while the other is the $-i\varepsilon$ path), the analytic continuation of $s_{\pm}$ is given by
\begin{align}
 s_{\pm} \to s_2+s_3 + \frac{-(s_3+m_1^2-m_2^2)(s_2+m_1^2-m_3^2) \pm \sqrt{ \lambda(s_2, m_1^2, m_3^2) \lambda(s_3, m_1^2, m_2^2)} }{2m_1^2}
 \,.
 \label{spmafter2}
\end{align}
As shown in Fig.~\ref{fig:generation} (right), $s_1=s_+$ is the singularity on the first sheet while $s_1=s_-$ may not be. Therefore, the position of the triangle singularity depends on whether the $s_2$ and $s_3$ are decaying modes $P$ or growing modes $P'$.

Let us see this difference in the equal mass case $m_1=m_2=m_3=m$. For $s_2=s_3=M^2$, the plus branch is given by
\begin{align}
s_+|_{s_2=s_3}=-\frac{M^2}{m^2}(M^2-4m^2)
\,,
\end{align}
whereas in the conjugate case $s_2=M^2, s_3=(M^2)^*$, it is given by
\begin{align}\label{eq: tTriang}
s_+|_{s_2=s_3^*}&=\frac{1}{2m^2}\left[ |M^2| \sqrt{(\re M^2-4m^2)^2+(\im M^2)^2} - \re M^2 (\re M^2 -4m^2) - (\im M^2)^2 \right]
\,.
\end{align}
The expressions are valid in both $\re M^2 < 4m^2$ and $\re M^2> 4m^2$. They agree with each other in the stable region $M^2 < 4m^2$ with $\im M^2 \to 0$. However, one can see
\begin{align}
s_+|_{s_2=s_3^*}&=\frac{4m^2 (\im M^2)^2 }{\re M^2(\re M^2 - 4m^2)}+\mathcal{O}((\im M^2)^4) 
\,,
\end{align}
in the unstable region $\re M^2 > 4m^2$ with the limit $\im M^2 \to 0$, which does not agree with $s_+|_{s_2=s_3}$ with the same limit $\im M^2 \to 0$.

Using the above analysis, we discuss analytic structures of 2-to-2 amplitudes by fixing the external states. For simplicity, we consider the system with one stable particle of the mass $m$ and one unstable particle of the complex mass $M$. We then consider the following amplitudes:
\begin{align}
\Amp_4^{++} &=
\begin{tikzpicture}[baseline=-2]
\begin{feynhand}
\vertex [particle] (1) at (-1,-0.6) {$1~$};
\vertex [particle] (2) at (-1,0.6) {$2^+$};
\vertex [particle] (3) at (1,0.6) {$3^+$};
\vertex [particle] (4) at (1,-0.6) {$4~$};
\vertex [grayblob] (a) at (0,0) {};
\propag  (4) -- (a) -- (1);
\propag [boson] (2) -- (a) -- (3);
\end{feynhand}
\end{tikzpicture}
\,, \qquad
\Amp_4^{+-}=
\begin{tikzpicture}[baseline=-2]
\begin{feynhand}
\vertex [particle] (1) at (-1,-0.6) {$1~$};
\vertex [particle] (2) at (-1,0.6) {$2^+$};
\vertex [particle] (3) at (1,0.6) {$3^-$};
\vertex [particle] (4) at (1,-0.6) {$4~$};
\vertex [grayblob] (a) at (0,0) {};
\propag  (4) -- (a) -- (1);
\propag [boson] (2) -- (a) -- (3);
\end{feynhand}
\end{tikzpicture}
\,, 
\label{stable_unstable}
\end{align}
where the solid and wavy lines represent stable and unstable particles, respectively. The amplitudes $\Amp_4^{+\pm}$ can have the following triangle singularities:
\begin{align}
\Amp_4^{+\pm} \sim
\underbrace{
\begin{tikzpicture}[baseline=-2]
\begin{feynhand}
\vertex [particle] (1) at (-1.25,-0.6) {};
\vertex [particle] (2) at (-1.25,0.6) {$+$};
\vertex [particle] (3) at (1.25,0.6) {$\pm$};
\vertex [particle] (4) at (1.25,-0.6) {};
\vertex (a) at (-0.4, -0.6) ;
\vertex (b) at (-0.4, 0.6) ;
\vertex (c) at (0.4, 0 ) ;
\propag (a) -- (b) -- (c) -- (a);
\propag            (a) -- (-1,-0.6);
\propag [boson] (b) -- (-1,0.6);
\propag [boson] (c) -- (1,0.6);
\propag           (c) -- (1,-0.6);
\end{feynhand}
\end{tikzpicture}
+
\begin{tikzpicture}[baseline=-2]
\begin{feynhand}
\vertex [particle] (1) at (-1.25,-0.6) {};
\vertex [particle] (2) at (-1.25,0.6) {$+$};
\vertex [particle] (3) at (1.25,0.6) {$\pm$};
\vertex [particle] (4) at (1.25,-0.6) {};
\vertex (a) at (0.4, -0.6) ;
\vertex (b) at (0.4, 0.6) ;
\vertex (c) at (-0.4, 0 ) ;
\propag (a) -- (b) -- (c) -- (a);
\propag            (c) -- (-1,-0.6);
\propag [boson] (c) -- (-1,0.6);
\propag [boson] (b) -- (1,0.6);
\propag           (a) -- (1,-0.6);
\end{feynhand}
\end{tikzpicture}
}_{s,u\text{-channels}} 
+
\underbrace{
\begin{tikzpicture}[baseline=-2]
\begin{feynhand}
\vertex [particle] (1) at (-1.25,-0.6) {};
\vertex [particle] (2) at (-1.25,0.6) {$+$};
\vertex [particle] (3) at (1.25,0.6) {$\pm$};
\vertex [particle] (4) at (1.25,-0.6) {};
\vertex (a) at (-0.6, 0.4) ;
\vertex (b) at (0.6, 0.4) ;
\vertex (c) at (0, -0.4 ) ;
\propag (a) -- (b) -- (c) -- (a);
\propag            (c) -- (-1,-0.6);
\propag [boson] (a) -- (-1,0.6);
\propag [boson] (b) -- (1,0.6);
\propag           (c) -- (1,-0.6);
\end{feynhand}
\end{tikzpicture}
}_{t\text{-channel}} ,
\end{align}
where the diagrams on the RHS show possible kinematic configurations of triangle singularities with specified external states. The all-$(+)$ amplitude $\Amp_4^{++}$ have triangle singularities at
\begin{align}
s,u&=m^2+\frac{1}{2}\left[ M^2 - \sqrt{3M^2(4m^2-M^2)}\right]
\,, \nonumber\\
t&= -\frac{M^2}{m^2}(M^2-4m^2)
\,,
\end{align}
whereas the positions of the triangle singularities of the mixed amplitude $\Amp_4^{+-}$ are
\begin{align}
s,u &= m^2+\frac{1}{2}\left[ M^2 - \sqrt{3M^2(4m^2-M^2)}\right], ~ m^2+\frac{1}{2}\left[ M^2 - \sqrt{3M^2(4m^2-M^2)}\right]^*
\,, 
\label{anomalous_su}
\end{align}
and the $t$-channel triangle is given in eq.(\ref{eq: tTriang}). To simplify the discussion, we assume a small decay width and approximate $M^2$ by a real number. Then, the $t$-channel triangle singularity of $\Amp_4^{++}$ exists at a negative $t$ region while that of $\Amp_4^{+-}$ is  $t\simeq 4m^2 (\im M^2)^2 /[\re M^2(\re M^2 - 4m^2)] \to +0$. Thus the triangle singularity for the $\Amp_4^{++}$ amplitude exits from the analytic continuation of the physical region, while the $\Amp_4^{+-}$ stems from the unphysical regime. This is precisely why we did not see the $t$-channel triangle singularity for  $\Amp_4^{+-}$ in our analysis from unitarity equations of stable particles in the previous section. Indeed information extracted from the unitarity equations is only applicable for physical kinematics, or analytic continuation thereof. Thus we see that contrary to the conclusion from the unitarity analysis, anomalous thresholds appear on the first sheet for both unstable particle scattering  $\Amp_4^{++}$  and  $\Amp_4^{+-}$.


\subsection{Anomalous thresholds for unstable particles: UV}
\label{sec:singularityUV}

In the previous discussion, we have assumed that the external kinematics are comparable to the internal masses, and thus the triangle is essentially an  ``IR''-loop, i.e. contributions that are computable within the IR theory and thus can be subtracted from any dispersive representation. We now move on to consider UV-loops where one or more heavy (unstable) particles are in the loop. These are not calculable from the IR theory, it is important to know when do the triangle singularities of these integrals enter the first sheet.  

In the event of a UV state in the loop, one can easily have $s_3< (m_1-m_2)^2$, where $m_2$ is the UV state. To track things, first consider the case  $s_3=(m_1-m_2)^2$ where the roots \eqref{s1roots} are degenerate $s_{\pm}=(m_2+m_3)^2+\frac{m_2}{m_1}[s_2-(m_1+m_3)^2]$ and outside of the integration region. The roots monotonically increase as $s_2$ increases and touch the endpoint $s_1'=(m_2+m_3)^2$ at the decay threshold $s_2=(m_1+m_3)^2$. Analytically continuing across the decay threshold brings the triangle singularity onto the first sheet. A similar analysis applies for the case where  $s_3< (m_1-m_2)^2$. In summary, for $s_3\leq (m_1-m_2)^2$, the UV triangle singularity enters the first sheet \textbf{only if} $s_2$ crosses the decay threshold! This is very different from the previous IR discussion, where $s_\pm$ can pinch the contour within a parameter range including \textbf{both} stable and unstable kinematics, i.e. eq.(\ref{Conditions}).

\begin{figure}[h]
\centering
\includegraphics[width=0.8\linewidth]{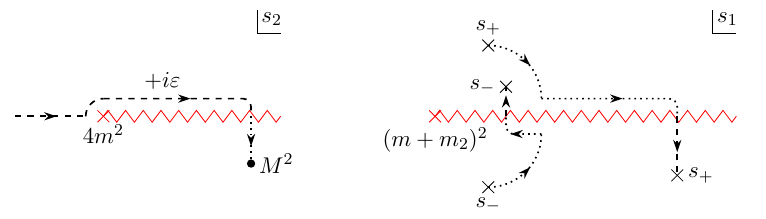}
\caption{The path of $s_2$ and the corresponding paths of $s_{\pm}(s_2)$ in the complex $s_1$ plane. As in Fig.~\ref{fig:generation}, the figures are drawn for illustrative purposes and the curves are not accurate.}
\label{fig:UVIRloop}
\end{figure}

Let us discuss the appearance of such singularities more concretely. For the sake of simplicity, we assign the same light particle to $m_1$, $m_3$, and $s_3$, and assume $m_2$ is much heavier than the light particle:
\begin{align}
m_2^2 \gg m^2= m_1^2=m_3^2=s_3
\,.
\end{align}
The positions of the triangle singularities in the $s_1$ plane are
\begin{align}
s_{\pm}&=m^2 + m_2^2 \frac{s_2}{2m^2} \pm \frac{\sqrt{m_2^2(m_2^2-4m^2) s_2(s_2-4m^2)}}{2m^2} 
\nn
&= m_2^2 \left[ \frac{s_2}{2m^2} \pm \frac{\sqrt{s_2(s_2-4m^2)}}{2m^2} \right]+\mathcal{O}\left(\frac{m^2}{m^2_2}\right)
.
\label{triangleUVIR}
\end{align}
The singularities are on the second sheet when $s_2$ is stable, $s_2<4m^2$. We then move $s_2$ along the path as shown in the left panel of Fig.~\ref{fig:UVIRloop}. The corresponding paths of $s_{\pm}(s_2)$ are shown in the right panel. The roots $s_{\pm}$ are complex during $s_2<4m^2$ and then collide and become degenerate at $s_2=4m^2$, for which $s_{\pm}=2m^2_2$ and thus far beyond the branch point $s_1=(m+m_2)^2$. We continue by adding a positive imaginary part to $s_2$. The square root remains the same form $\sqrt{s_2(s_2-4m^2)} \to \sqrt{s_2(s_2-4m^2)}$ around $s_2=4m^2$ by using the principal square root, so $s_{\pm}$ moves to the right/left with a positive/negative imaginary part as $\re\, s_2$ increases. Then, as $s_2$ is continued to the lower-half plane, $s_{\pm}$ passes through the real axis and shows up on the first sheet of the complex $s_1$ plane.

\begin{figure}[h]
\centering
\includegraphics[width=0.8\linewidth]{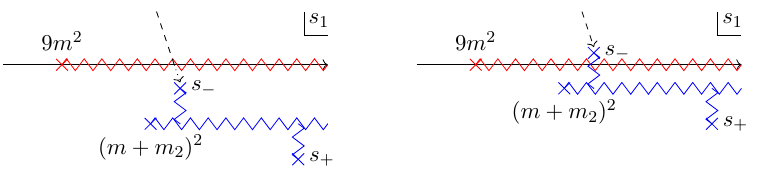}
\caption{The expected positions of the triangle singularities from the loop with a UV state. Left: the triangle singularities are in the unphysical sheet when $m_2$ has a sufficiently large decay width. Right: the triangle singularities can show up on the physical sheet if the heavy particle is long-lived.}
\label{fig:UVIRamp}
\end{figure}

One may suspect that the appearance of such singularities is an artifact of our one-loop truncation. In principle,  the UV mass $m_2$ should be complexified and the branch cut running from $(m+m_2)^2$ is not a normal-threshold cut on the real axes, but on the complex plane on an unphysical sheet~\cite{Zwanziger:1963zza, Lvy1959OnTD, PhysRev.119.1121, PhysRev.123.692, Landshoff:1963nzy, Eden:1966dnq}. In particular, as the decay of $m_2$ particle is a 2-loop effect, it is at the same order as the the three-particle threshold. In such case, whether the triangle singularities appear on the physical sheet or not depends on the decay widths of the external unstable particle and the internal heavy particle as illustrated in Fig.~\ref{fig:UVIRamp}.

While we need a higher-loop analysis for precise treatment, we suppose that the positions of the triangle singularities are still given by \eqref{triangleUVIR} but now with complexified $m_2^2$. We discuss when $s_-$ shows up on the physical sheet where $s_-$ is the singularity appearing above the complex normal threshold. For simplicity, we assume $m^2 \ll |M^2| (\ll |m_2^2|)$ and find
\begin{align}
s^{(+i\varepsilon)}_- &\simeq m_2^2\left(1+ \frac{m^2}{M^2} \right)
\,, \nonumber\\
s^{(-i\varepsilon)}_- &\simeq m_2^2 \frac{(M^2)^*}{m^2}\,, 
\end{align}
where the superscripts denote which path we chose. The imaginary parts are
\begin{align}
\im\, s^{(+i\varepsilon)}_- &\simeq m_R \left( \frac{m^2 m_R}{M_R^3}\Gamma - \gamma \right)
\,, \nonumber\\
\im\, s^{(-i\varepsilon)}_- &\simeq \frac{m_R M_R}{m^2} \left( m_R \Gamma - M_R \gamma \right)
\,,
\end{align}
where $m_2^2=m_R^2-i m_R \gamma$ and $M^2=M_R^2-i M_R \Gamma$ with $0<\gamma \ll m_R$ and $0<\Gamma \ll M_R$. Therefore, the triangle singularity from the loop with a UV state may appear on the upper-half plane when
\begin{align}
\gamma < \frac{m^2 m_R}{M_R^3} \Gamma \qquad &\text{for the $+i\varepsilon$ path,} \nonumber\\
\gamma < \frac{m_R}{M_R} \Gamma \qquad  &\text{for the $-i\varepsilon$ path.}
\end{align}
Hence, the analytic structure is similar to the one-loop as long as the decay of the UV state is sufficiently small.




\section{A violation of positivity bounds: toy example}
\label{sec:example}

In non-gravitational EFTs, one can in general derive positivity bounds for the four-derivative operator based on the optical theorem for the $a,b\rightarrow a,b$ four-point amplitude, where $a,b$ labels the potential distinct particle species. As discussed in the previous sections, when there are unstable particles, we have anomalous thresholds from the UV which are not subtractable and thus contribute to the dispersive representation. Such contributions would violate the positivity bounds. In this section, we consider a simple toy example to illustrate such a phenomenon.

Consider a theory with four scalars, $\pi, \phi, \chi_L$ and $\chi_H$, where the mass scales are such that $\pi$ is the lightest state and consider the following interactions:
\begin{align}
\mathcal{L}_{\rm int}= - \frac{g_{\phi}}{2} \phi \chi_L^2 - g_{\pi} \pi \chi_L \chi_H
\,.
\end{align}
As a result the leading contribution to the four-point $\Amp \left(\pi(1) \phi(2) \phi(3) \pi(4) \right)$ is the sum of the scalar box integral:
\begin{align}
\Amp(s,t; s_2,s_3)&=
\begin{tikzpicture}[baseline=0]
\begin{feynhand}
\propag (-1.2, -0.8) node [left] {$p_1$} -- (-0.6, -0.6) ;
\propag [photon] (-1.2, 0.8) node [left] {$p_2$} -- (-0.6, 0.6) ;
\propag [photon] (1.2, 0.8) node [right] {$p_3$} -- (0.6, 0.6) ;
\propag (1.2, -0.8) node [right] {$p_4$} -- (0.6, -0.6) ;
\propag [dashed] (-0.6, -0.6) -- (-0.6, 0.6) node [midway, left=0.05] {$m_L$};
\propag [dashed]  (-0.6, 0.6) -- (0.6, 0.6) node [midway, above=0.05] {$m_L$};
\propag [dashed]  (0.6, 0.6) -- (0.6, -0.6) node [midway, right=0.05] {$m_L$};
\propag [ultra thick]  (0.6, -0.6) -- (-0.6, -0.6) node [midway, above=0.05] {$m_H$};
\end{feynhand}
\end{tikzpicture}
+
\begin{tikzpicture}[baseline=0]
\begin{feynhand}
\propag (-1.2, -0.8) node [left] {$p_1$} -- (0.6, -0.6) ;
\propag [photon] (-1.2, 0.8) node [left] {$p_2$} -- (-0.6, 0.6) ;
\propag [photon] (1.2, 0.8) node [right] {$p_3$} -- (0.6, 0.6) ;
\propag (1.2, -0.8) node [right] {$p_4$} -- (-0.6, -0.6) ;
\propag [dashed] (-0.6, -0.6) -- (-0.6, 0.6) node [midway, left=0.05] {$m_L$};
\propag [dashed]  (-0.6, 0.6) -- (0.6, 0.6) node [midway, above=0.05] {$m_L$};
\propag [dashed]  (0.6, 0.6) -- (0.6, -0.6) node [midway, right=0.05] {$m_L$};
\propag [ultra thick]  (0.6, -0.6) -- (-0.6, -0.6) node [midway, above=0.05] {$m_H$};
\end{feynhand}
\end{tikzpicture}
\nn
&=\frac{g_{\phi}^2 g_{\pi}^2}{(4\pi)^2}\left[ \mathcal{I}_{\rm box}(s, t; s_2, s_3) + \mathcal{I}_{\rm box}(u, t; s_2, s_3) \right] 
\,, \\
\mathcal{I}_{\rm box}(s,t;s_2,s_3) &= \int \frac{\D^4 \ell}{i\pi^2}\frac{1}{(q_1^2+m_L^2)(q_2^2+m_L^2)(q_3^2+m_L^2)(\ell^2+m_H^2)}
\,.
\label{M_example}
\end{align}
Here, $s=-(p_1+p_2)^2, t=-(p_1+p_4)^2, u=-(p_1+p_3)^2, s_i=-p_i^2$ and $q_i$ are
\begin{align}
q_1=\ell+p_1 \,, \quad
q_2=\ell +p_1 + p_2 \,, \quad
q_3=\ell+p_1+p_2+p_3 \,,
\end{align}
where $\phi, \pi, \chi_L$ and $\chi_H$ correspond to the wavy, solid, dashed, and thick lines in the Feynman diagrams, respectively. The particle $\chi_H$ is supposed to be heavy so that we can separate the physics of the IR particles $(\phi, \pi, \chi_L)$ and the UV particle $\chi_H$. The low-energy part of $\Amp$ can be described by a low-energy EFT composed of the particles $(\phi, \pi, \chi_L)$. We shall discuss how the low-energy part of $\Amp$ changes when we change the mass spectrum of the IR particles $(\phi, \pi, \chi_L)$. The masses of $(\phi, \pi, \chi_L)$ are denoted by $(M, m, m_L)$, respectively.

When the decay $\phi \to \chi_L \chi_L$ is kinematically allowed, the mass of $\phi$ is complexified. We use $M_R^2$ to denote the real part of $M^2$. The leading order result of the decay width is
\begin{align}
\Gamma =\frac{g_{\phi}^2}{32\pi M_R} \sqrt{1-4m_L^2/M_R^2} \, \theta(M_R^2-4m_L^2)
\,,
\end{align}
and then the mass squared is
\begin{align}
M^2=M_R^2 -i M_R \Gamma
\,,
\end{align}
where $\theta(x)$ is the step function. To deal with the external unstable particles, we first regard $s_2$ and $s_3$ as complex variables and analytically continue the box integrals. On the other hand, we set $s_2=s_3=M_R^2$ for the stable kinematic of $\phi~(M_R<2m_L)$. Note that the decay width for $\chi_H \rightarrow \pi, \chi_L$ pair is controlled by $g_{\pi}$, which can be taken to be small in a controlled fashion and thus neglected in the remaining discussion. 

For a small fixed $t$, $|t|\ll 4m_L^2$, the amplitude $\Amp(s)$ must be analytic in a small-$|s|$ region because the $s$ and $u$ channel cuts run from the heavy mass scale $m_H^2$ (see also below). The Taylor expansion around the $s\leftrightarrow u$ symmetric point is then
\begin{align}
\Amp=B_0(t; s_2, s_3) + B_2(t; s_2, s_3) [s-m^2-(s_2+s_3)/2+t/2]^2 + \mathcal{O}(s^4)
\,.
\end{align}
The $s^2$ coefficient $B_2$ can be directly read off by the low-energy expansion of the amplitude. On the other hand, as is well-known, when $\phi$ is stable and no anomalous threshold exists, the low-energy coefficients admit representations by the use of the high-energy integral 
\begin{align}
B_2=\int_{(m_L+m_H)^2}^{\infty} \frac{\D s}{2\pi i} \frac{2\Disc_s \Amp}{[s-m^2-(s_2+s_3)/2+t/2]^{3}}
\,.
\label{sum_rule_UVIR}
\end{align}
Unitarity ensures the positivity of the discontinuity $\im\, \Amp = \Disc_s \Amp/2i>0$ in the forward limit $t\to 0$. We thus find the inequality $B_2|_{t=0}>0$ which is known as the positivity bound~\cite{Adams:2006sv}. 

One may expect that the same inequality holds even in the unstable $\phi$ particle if we choose the conjugate pair $s_2=s_3^*=M^2$. Unitarity then leads to the positivity of the discontinuity with $s_2=s_3^*=M^2$ at the forward limit $t=t_0$. We can practically take $t_0 \to 0$ because $t_0$ scales as $t_0\simeq 4m^2(\im M^2)^2/s^2$ with $s>m_H^2$. So we end up with $B_2|_{t=0}>0$ {\it if the dispersion relation \eqref{sum_rule_UVIR} remains to be true in unstable particles}. As we've seen in the previous discussion, the presence of anomalous thresholds can potentially spoil eq.(\ref{sum_rule_UVIR}). For stable particles, due to the UV scale of $m_H^2$ the triangle singularity is never on the first sheet, as can be seen in eq.(\ref{Conditions}). For unstable particles, triangle singularities in the UV can contribute, and thus $B_2$ might become negative.

Consider the low-energy expansion of the box integral
\begin{align}
\mathcal{I}_{\rm box}(s,t;s_2,s_3)=\int \frac{\D^4 \ell}{i\pi^2}\frac{1}{(q_1^2+m_L^2)(q_2^2+m_L^2)(q_3^2+m_L^2)(\ell^2+m_H^2)}
\end{align}
under the hierarchy $m_L,\, |p_i| \ll m_H$. We wish to compute the coefficient for $s^2$ in the low energy expansion. To track the expansion more concretely, we separate the loop momentum into two regions based on an intermediate scale $\Lambda$ such that $m_L,\, |p_i| \ll \Lambda \ll m_H$ and divide the integral into small and large regions~\cite{Smirnov:2002pj}:
\begin{align}
\mathcal{I}_{\rm box} = \mathcal{I}_{\rm small}+ \mathcal{I}_{\rm large}=\left( \int_{|\ell|<\Lambda} {+}  \int_{|\ell|>\Lambda}\right) \frac{1}{(q_1^2+m_L^2)(q_2^2+m_L^2)(q_3^2+m_L^2)(\ell^2+m_H^2)}
\,, \label{divide_box}
\end{align}
For $\mathcal{I}_{\rm small}$, all kinematics are parametrically smaller than $m_H$ and one can Taylor expand the integrand in $1/m_H^2$. For $\mathcal{I}_{\rm large}$ since $\ell\gg m_L,\, |p_i|$ we can expand the propagators that does not contain $m_H$ in $1/\ell^2$. This amounts to the following expansion in the two regions respectively, 
\begin{align}
\mathcal{I}_{\rm small}: \quad& \frac{1}{\ell^2+m_H^2}= \frac{1}{m_H^2} - \frac{\ell^2}{m_H^4} + \frac{(\ell^2)^2}{m_H^6} + \cdots\,,
\nonumber\\
\mathcal{I}_{\rm large}:  \quad&\frac{1}{(\ell + P_i) +m_L^2} = \frac{1}{\ell^2} - \frac{2\ell \cdot P_i +P_i^2 + m_L^2}{(\ell^2)^2} + \frac{(2\ell \cdot P_i +P_i^2 + m_L^2)^2}{(\ell^2)^3} +\cdots\,,
\end{align}
with $P_i=\sum_{j=1}^i p_j$. For $\mathcal{I}_{\rm small}$ we will be working with the $t$-channel triangle integral with loop momentum dependent numerators:
\begin{align}\label{IntTriangle}
\frac{1}{m_H^{2n{+}2}} \int \frac{\D^4 \ell}{i\pi^2} \frac{(\ell^2)^n}{(q_1^2+m_L^2)(q_2^2+m_L^2)(q_3^2+m_L^2)}
\,.
\end{align}
Since after integral reduction, this only generates $t$-channel scalar triangle and bubbles, $s$-dependence can only come from the numerator in the process of reduction. Since we are interested in the $s^2$ coefficient, only $n=2$ is relevant, giving a coefficient that scales as $\mathcal{O}(m_H^{-6})$. On the other hand for $\mathcal{I}_{\rm large}$, since 
\begin{align}
\int \D^4 \ell \frac{(2\ell\cdot P)^j}{(\ell^2)^i (\ell^2+m_H^2)} &\propto 
\begin{cases}
0 &{\rm for}~~j={\rm odd}\\
\frac{(P^2)^{j/2}}{m_H^{2(i-1-j/2)}}
&{\rm for}~~j={\rm even}
\,,
\end{cases}
\end{align}
we see that contributions to the $s^2$ coefficient starts at $\mathcal{O}(m_H^{-8})$. Thus in conclusion, $\mathcal{I}_{\rm small}$ yields the leading order contribution to $B_2$ in the large $m_H^{2}$ expansion.

Integrating eq.(\ref{IntTriangle}) with $n=2$, we find
\begin{align}
B_{2}&=\frac{g_{\phi}^2g_{\pi}^2}{(4\pi)^2m_H^6 \hat{\lambda}^2}
\nn
&\quad \times
\Biggl[ \left\{(s_2-s_3)\hat{\lambda}+3t (s_3-t)^2-3t s_2^2 \right\} \Lambda(s_2)
+ \left\{(s_3-s_2)\hat{\lambda}+3t (s_2-t)^2-3t s_3^2 \right\} \Lambda(s_3)
\nn
&\qquad +6t^2(s_2+s_3-t)\Lambda(t) + 2t\left\{ (2m_L^2+t)\hat{\lambda}+6t s_2 s_3 \right\} \mathcal{I}_{\rm tri}(t, s_2, s_3) - 2 t\hat{\lambda} 
\Biggl]
+\mathcal{O}(m_H^{-8})
\label{B2leading}
\end{align}
where $\hat{\lambda}:=\lambda(s_2,s_3,t)=(s_2-s_3)^2+\mathcal{O}(t)$, $\mathcal{I}_{\rm tri}$ is the triangle integral with all internal masses being $m_L$, and
\begin{align}
\Lambda(z)=\sqrt{1-\frac{4m_L^2}{z}} \ln\left[ -\frac{1-\sqrt{1-\frac{4m_L^2}{z}}}{1+\sqrt{1-\frac{4m_L^2}{z}}}\right]
\,,
\end{align}
is the discontinuous part of the bubble integral; $\Lambda(z)$ has a branch cut along $z>4m_L^2$ and analytic elsewhere on the first sheet.

We would like to evaluate $B_{2}$ at $s_2=s_3^*=M^2$ and $t\to +0$. However, we should carefully take the limit $t \to +0$ since $\mathcal{I}_{\rm tri}$ can be singular at $t=0$ when $s_2$ and $s_3$ are analytically continued to the second sheet.  Nevertheless, we can make sure $\lim_{t\to +0}t \mathcal{I}_{\rm tri} =0$ independently from the details of analytic continuation as follows. For general kinematics, the triangle integral is expressed by 12 Spence's functions~\cite{tHooft:1978jhc, Patel:2015tea, Patel:2016fam}:
\begin{align}
\mathcal{I}_{\rm tri}=\frac{1}{\lambda^{1/2}(s_2,s_3,t)}\sum_{a=\pm 1}\left[ {\rm Li}_2 \xi^{+1}_a(s_2,s_3,t) -{\rm Li}_2 \xi^{-1}_a(s_2,s_3,t) \right] + (s_2\leftrightarrow s_3) + (s_2 \leftrightarrow t)\,,
\end{align}
with
\begin{align}
\xi^b_a(s_2,s_3,t)=\frac{s_2(s_2-s_3-t)+b s_2 \sqrt{\lambda(s_2,s_3,t)}}{s_2(s_2-s_3-t)+a \sqrt{\lambda(s_2,m_L^2,m_L^2)\lambda(s_2,s_3,t)}}
\,.
\end{align}
Spence's function ${\rm Li}_2(z)$ has a logarithmic branch point at $z=1$ and we have to choose an appropriate branch to evaluate the concrete value of $\mathcal{I}_{\rm tri}$. In the present purpose, on the other hand, we do not choose the branch and use the expression on a general branch
\begin{align}
{\rm Li}_2(z)={\rm pv \, Li}_2 (z) + 2 \pi n i \ln z + 4\pi^2 m \quad (n,m=0, \pm 1, \pm 2,\cdots)
\,,
\end{align}
where ${\rm pv}$ denotes the principal value. It is easy to show that $\mathcal{I}_{\rm tri}$ can diverge as $t\to 0$ on a general branch but is a logarithmic divergence. Therefore, we conclude $\lim_{t\to +0}t \mathcal{I}_{\rm tri} =0$ and ignore the second line of \eqref{B2leading} in the limit $t\to +0$ as long as $s_2 \neq s_3$. We then obtain the following expressions at the leading order in the large $m_H$ expansion:
\begin{align}
B_{2}\approx
\frac{g_{\phi}^2 g_{\pi}^2}{(4\pi)^2 m_H^6} \times
\begin{cases}
-\frac{1}{s_2}-\frac{2m_L^2 \Lambda(s_2)}{s_2(4m_L^2-s_2)}&(s_2=s_3)
\,,
 \\
 \frac{\Lambda(s_2)-\Lambda(s_3)}{s_2-s_3} &(s_2\neq s_3)
\,,
\end{cases}
\label{B2t=0}
\end{align}
where ``$\approx$'' is used to denote the equality at the leading order in the large mass expansion of $m_H$. For the sake of comparison, we computed $B_{2}$ in which $s_2=s_3$ and $t=0$ are assumed on the physical sheet. The former one of \eqref{B2t=0} agrees with the latter one with the limit $s_3 \to s_2$. 

\begin{figure}[t]
\centering
 \includegraphics[width=0.6\linewidth]{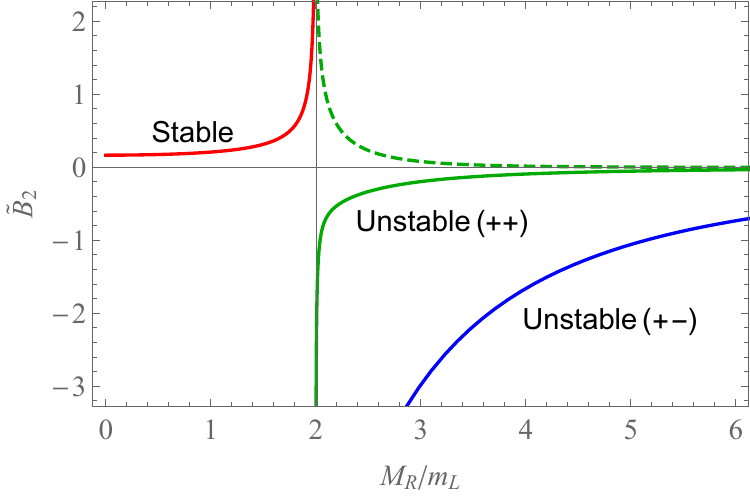}	
\caption{Normalized $s^2$ coefficient as $\tilde{B}_{2}=\frac{(4\pi)^2m_H^6}{g_{\phi}^2g_{\pi}^2 m_L^2} B_{2}$: stable external particles $B_{2}^{\rm stable}$ (red), decaying-decaying external particles $B_{2}^{++}$ (green), and decaying-growing external particles $B_{2}^{+-}$ (blue). While $B_{2}^{\rm stable}$ and $B_{2}^{+-}$ are real numbers, $B_{2}^{++}$ is a complex number; the real and imaginary parts are represented by solid and dashed curves, respectively. We set $g_{\phi}^2/4\pi = M_R^2$ for illustrative purpose. Note that the divergence at $M_R=2m_L$ is caused by a $t$-channel triangle singularity of $\Amp$. }
\label{figB2}
\end{figure}

We now discuss $B_{2}$ in the stable case $(s_2=s_3<4m_L^2)$, the unstable case with the same choice $(s_2=s_3=M^2)$, and the unstable case with the conjugate choice $(s_2=s_3^*=M^2)$, respectively. The function $\Lambda(z)$ has a branch cut in $z>4m_L^2$ and the analytically continued $\Lambda$ on the unphysical sheet is given by
\begin{align}
\Lambda^{\pm} (z)&={\rm pv}\, \Lambda(z) \pm 2\pi i \frac{\sqrt{z(z-4m_L^2)}}{z} 
&(\re z>4m_L^2,~{\rm sgn}( \im z )= \mp)
\,,
\label{LambdaAC}
\end{align}
with $\Lambda^-(z^*)=[\Lambda^+(z)]^*$. As a result, we obtain
\begin{align}
B_{2}^{\rm stable}&\approx -\frac{g_{\phi}^2 g_{\pi}^2}{(4\pi)^2 m_H^6} \left[ \frac{1}{M_R^2} + \frac{2m_L^2 \Lambda(M_R^2)}{M_R^2(4m_L^2-M_R^2)} \right] &(M_R<2m_L)
\,,\nonumber\\
B_{2}^{++}&\approx -\frac{g_{\phi}^2 g_{\pi}^2}{(4\pi)^2 m_H^6} \left[ \frac{1}{M^2} + \frac{2m_L^2 \Lambda^+ (M^2)}{M^2(4m_L^2-M^2)} \right]
&(M_R>2m_L)
\,, \nonumber\\
B_{2}^{+-}&\approx  - \frac{g_{\phi}^2 g_{\pi}^2}{(4\pi)^2 m_H^6} \frac{\im \Lambda^+(M^2) }{M_R \Gamma}
&(M_R>2m_L)
\,.
\label{Results}
\end{align}
 The sign symbol denotes whether the unstable particle is decaying $(+)$ or growing $(-)$. The behaviours of $B_{2}$ are shown in Fig.~\ref{figB2}. Thus we see in this explicit toy model, the standard positivity bound $B_2>0$ is violated for the unstable particles. Again this result is not surprising given for unstable particles, anomalous thresholds from UV massive state do contribute. In the next section, we will show that the source of the negativity indeed arises from the anomalous threshold. As we have mentioned, the violation of $B_2>0$ would imply that the dispersion relation \eqref{sum_rule_UVIR} no longer holds in unstable particles.

\section{Anomalous thresholds from double discontinuity}
\label{sec:dispersion_relation}
In this section, we will find a dispersion relation of $B_2^{+-}$ with a special emphasis on delineating contributions from the normal and anomalous thresholds, and see that the anomalous thresholds are indeed the origin of the negative sign.

\begin{figure}[h]
\centering
\adjincludegraphics[valign=c, scale=0.5]{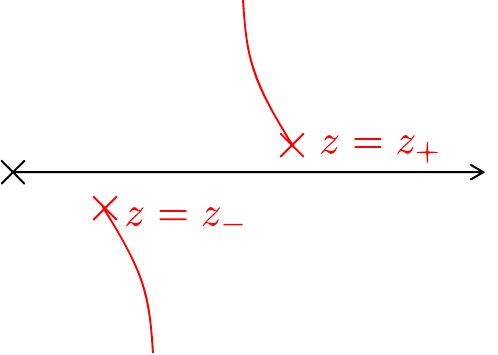}
~
$\to$
~
\adjincludegraphics[valign=c, scale=0.5]{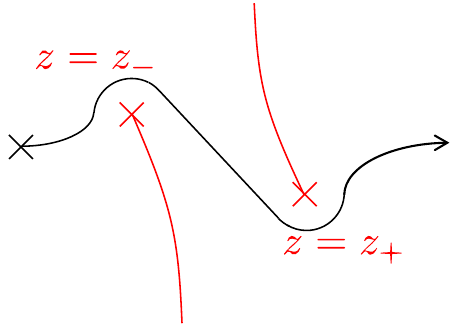}
~
$=$
~
\adjincludegraphics[valign=c, scale=0.5]{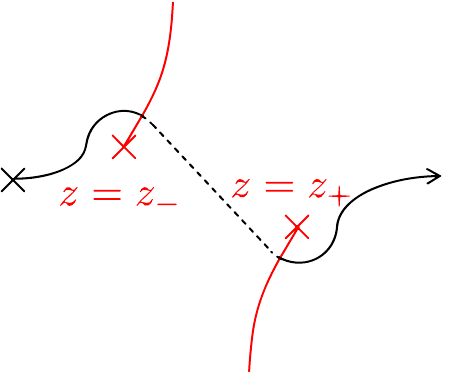}
\caption{Integration contours before (left) and after (middle and right) the analytic continuation. The red curves represent the branch cuts of the logarithmic function of \eqref{B2dispersion}. The black solid curves are the contours on the first sheet while the black dashed curve is the path on the second sheet. The branch cuts are deformed in the middle and right figures; in the right panel, the first sheet agrees with the principal sheet of the logarithm while it does not in the middle panel. The contours always run above $z=z_-$ and below $z=z_+$. }
\label{fig:contours}
\end{figure}

We start with the standard dispersive representation in the stable kinematics $(0<s_2,s_3<4m_L^2)$: the low-energy coefficient of the amplitude is given by the integral along the normal threshold cuts as
\begin{align}
\tilde{B}_2 &= \frac{m_H^6}{m_L^2} \int_{m_{\rm th}^2}^{\infty} \D s \frac{2\rho_{\rm box}|_{t=0}}{[s-m^2-(s_2+s_3)/2]^3}
\nn
&\approx
2\int^{\infty}_1 \D z  \frac{ m_L^2 }{s_3-s_2}  \frac{\ln[z-z_+(s_2)][z-z_-(s_2)]}{z^3} + (s_2 \leftrightarrow s_3)
\label{B2dispersion}
\end{align}
where $\tilde{B}_2 = \frac{(4\pi)^2 m_H^6}{g_{\phi}^2 g_{\pi}^2m_L^2} B_2, z=s/m^2_{\rm th}, m_{\rm th}^2=(m_L+m_H)^2$ is the threshold energy and
\begin{align}
z_{\pm}(s)=\frac{1}{2m_L^2}\left[s \pm \sqrt{s(s-4m_L^2)} \right]
\,,
\end{align}
are (the rescaled values of) the positions of the triangle singularities studied in Sec.~\ref{sec:singularityUV}. The integrand has singularities at $z=z_{\pm}(s_2)$ and $z=z_{\pm}(s_3)$ which are away from the real $z$-axis when $0<s_2,s_3<4m_L^2$ [see Fig.~\ref{fig:UVIRloop} and Fig.~\ref{fig:contours} (left)].

We then increase $s_2$ along the $+i\varepsilon$ path and $s_3$ along the $-i\varepsilon$ path respectively. Since the logarithm in the integrand of \eqref{B2dispersion} can be separated into pure $s_2$- and pure $s_3$-dependent pieces, we can discuss their analytic continuation separately. As long as $\im \, s_2>0$, i.e., $s_2$ does not cross the real axis, the singularities $z=z_{\pm}(s_2)$ do not touch the integration contour and the contour deformation is not needed. Eq.~\eqref{B2dispersion} is simply given by the integral of the principal value of the logarithm (multiplied by $z^{-3}$). Again, as $s_2=4m_L^2$ we enter unstable kinematics and we continue to the lower-half  plane as in the left of Fig.~\ref{fig:UVIRloop}, the singularities $z=z_{\pm}(s_2)$ cross the real axis and then the contour needs to be distorted (the middle of Fig.~\ref{fig:contours}). The contour remains to run above $z=z_-$ and below $z=z_+$. Note that since $\im \, s_2<0$, the branch cut for the principle branch of logarithm should run upward from $z=z_-$ and downward from $z=z_+$, as the integrand with $\im \, s_2<0$ is conjugate to that with $\im\, s_2>0$. We can redefine the branch cuts so that they align with the principal branch. The price we pay is that the contour will now pass through one branch cut onto the second sheet and come back through the second branch cut, as shown in the right of Fig.~\ref{fig:contours}. Thus we see that the analytically-continued $\tilde{B}_2$ is no longer given by a contour integral of the principal value of the logarithm.

It is convenient to divide the contour integral as follows:
\begin{align}
\adjincludegraphics[valign=c, scale=0.5]{contour2}
=
~
\adjincludegraphics[valign=c, scale=0.5]{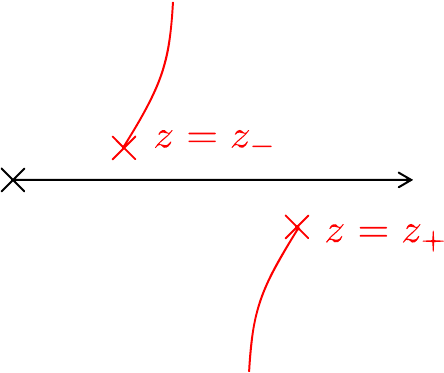}
+~
\adjincludegraphics[valign=c, scale=0.5]{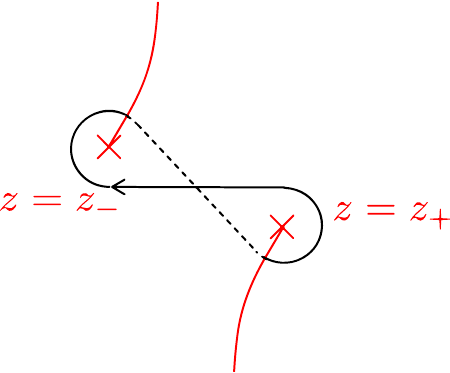}
\,.
\label{contour_deformation}
\end{align}
The first term is nothing but the integral of the principal logarithm while the second term is the integral of the difference between the different sheets, namely the discontinuity. Therefore, the analytically-continued dispersive representation is 
\begin{align}
\tilde{B}_2
\approx 
2\int^{\infty}_1 \D z  \frac{ m_L^2 }{s_3-s_2}  \frac{\ln[z-z_+(s_2)][z-z_-(s_2)]}{z^3} 
+2\int_{z_+(s_2)}^{z_-(s_2)} \D z \frac{ m_L^2 }{s_3-s_2}  \frac{2\pi i}{z^3}  + (s_2 \leftrightarrow s_3)
\,,
\label{B2dispersionAC}
\end{align}
where $\ln$ is understood as the principal logarithm. The integrations yield
\begin{align}
2\int^{\infty}_1 \D z \frac{ m_L^2 }{s_3-s_2}\frac{\ln[z-z_+(s_2)][z-z_-(s_2)]}{z^3}  &= \frac{{\rm pv}\, \Lambda(s_2)}{s_2-s_3} + \frac{1}{s_2-s_3}
\,, \\
2\int_{z_+(s_2)}^{z_-(s_2)} \D z \frac{ m_L^2 }{s_3-s_2}  \frac{2\pi i}{z^3} &= 2\pi i\frac{\sqrt{s_2(s_2-4m_L^2)}}{s_2(s_2-s_3)}
\,.
\end{align}
The second term of the first integral is cancelled with the $(s_2 \leftrightarrow s_3)$ term in \eqref{B2dispersionAC}. Hence, \eqref{B2dispersionAC} indeed reproduces \eqref{Results} by setting $s_2=s_3^*=M^2$.

All the calculations have been so far made explicit but the essential ingredients are (i) pairs of new singularities enter the first sheet and (ii) the integration contour is deformed like \eqref{contour_deformation}. The second term of \eqref{contour_deformation} is the evaluation of the discontinuity of the integrand. Since the later is the discontinuity of the amplitude in $s_1$, the second term is a double discontinuity. Hence, the dispersive representation of the $s^2$ coefficient at $t=0$ takes a more illustrative form:
\begin{align}
\tilde{B}_2^{+-}=\int_{m_{\rm th}^2}^{\infty} \frac{\D s}{2\pi i} \frac{2 \Disc_s \tilde{\Amp}^{+-} }{(s-\re M^2 -m^2)^3}
+ \sum_n \int_{\mathcal{C}_n} \frac{\D s}{2\pi i} \frac{2 \Disc_s^2 \tilde{\Amp}^{+-} }{(s-\re M^2 -m^2)^3}
\,.
\label{B2dispersionA}
\end{align}
Here, $\Disc_s^2 \Amp^{+-}$ refers to the double discontinuity of the same channel. The contour $\mathcal{C}_n$ is a path connecting the pair of singularities and the summation is over all the pairs.\footnote{If $\Disc_s^2 \Amp^{+-}$ has singularities and they require to deform the contour $\mathcal{C}_n$, we can again separate the integral into that of $\Disc_s^2 \Amp^{+-}$ and $\Disc_s^3 \Amp^{+-}$, which may continue to even higher orders. In our one-loop example, however, it is enough to include up to the double discontinuity.} In the example above, the discontinuity $\Disc_s \tilde{\Amp}^{+-}$ has two pairs of the triangle singularities associated with the contact diagrams
\begin{align*}
\begin{tikzpicture}[baseline=-2]
\begin{feynhand}
\propag[boson] (0.6,1)  -- (0, 0) ;
\propag (0.6,-1) --  (0, 0)  ;
\propag (-2, -1)  -- (-1.2, -0.6);
\propag[ultra thick] (-1.2, -0.6) -- (0, 0) ;
\propag[boson] (-2, 1) -- (-1.2 , 0.6) ;
\propag[dashed] (-1.2, 0.6) -- (0, 0);
\propag[dashed] (-1.2,-0.6) -- (-1.2,0.6)  ;
\end{feynhand}
\end{tikzpicture}
\qquad \text{and} \qquad
\begin{tikzpicture}[baseline=-2]
\begin{feynhand}
\propag[boson] (-0.6,1)  -- (0, 0) ;
\propag (-0.6,-1) --  (0, 0)  ;
\propag (2, -1)  -- (1.2, -0.6);
\propag[ultra thick] (1.2, -0.6) -- (0, 0) ;
\propag[boson] (2, 1) -- (1.2 , 0.6) ;
\propag[dashed] (1.2, 0.6) -- (0, 0);
\propag[dashed] (1.2,-0.6) -- (1.2,0.6)  ;
\end{feynhand}
\end{tikzpicture}
\,.
\end{align*}
Since the second term in \eqref{B2dispersionA} occurs only when the triangle singularity crosses onto the physical sheet, we can identify it with the anomalous threshold. Thus the normal and anomalous contributions have been identified with
\begin{align}
\tilde{I}_n^{+-} &:= \int_{m_{\rm th}^2}^{\infty} \frac{\D s}{2\pi i} \frac{2 \Disc_s \tilde{\Amp}^{+-} }{(s-\re M^2 -m^2)^3} 
\approx 4\im \int^{\infty}_1 \D z \frac{ m_L^2 }{2M_R \Gamma}\frac{\ln[z-z_+(M^2)][z-z_-(M^2)]}{z^3} 
\,, \nonumber\\
\tilde{I}_a^{+-} &:=\sum_n \int_{\mathcal{C}_n} \frac{\D s}{2\pi i} \frac{2 \Disc_s^2 \tilde{\Amp}^{+-}}{(s-\re M^2 -m^2)^3}
\approx 
4\re \int_{z_+(M^2)}^{z_-(M^2)} \D z \frac{ m_L^2 }{M_R \Gamma}  \frac{\pi }{z^3} 
\,.
\end{align}

\begin{figure}[t]
\centering
 \includegraphics[width=0.6\linewidth]{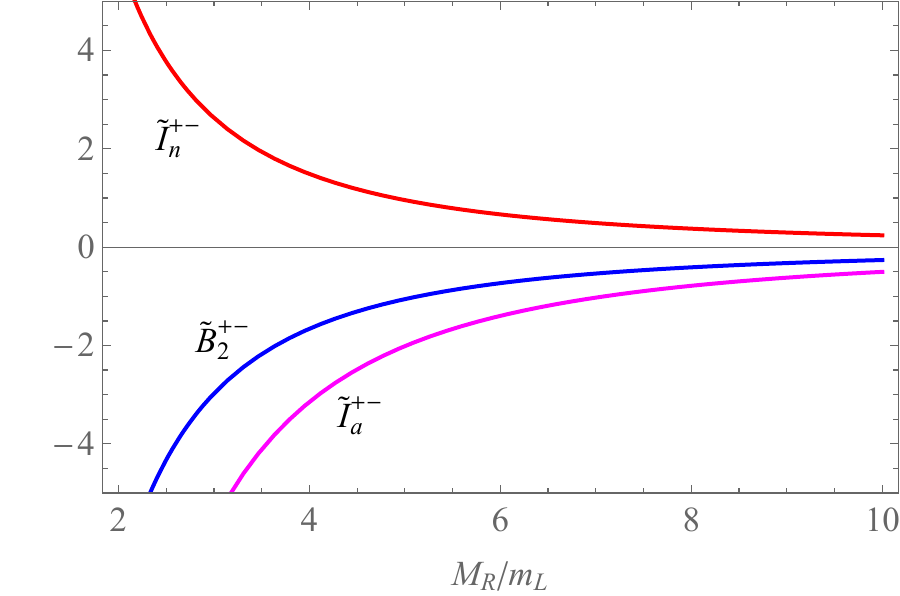}	
\caption{Normalized $s^2$ coefficient $\tilde{B}_2^{+-}$ (blue), the normal threshold contribution $\tilde{I}_n^{+-}$ (red), and the anomalous threshold contribution $\tilde{I}_a^{+-}$ (magenta).}
\label{figB2I}
\end{figure}

The dispersion relation \eqref{B2dispersionA} disentangles the contributions from normal and anomalous thresholds to $\tilde{B}_2^{+-}$. We have numerically computed their individual contribution  as in  Fig.~\ref{figB2I}.  As one can see the normal threshold $\tilde{I}_n^{+-}$ still satisfies the positivity $\Disc_s \tilde{\Amp}^{+-}/(2i)>0$ in the unstable kinematics $(M_R>2m_L)$ as predicted by unitarity, while the anomalous part $\tilde{I}_a^{+-}$ yields a large negative value. The anomalous sign of the $s^2$ coefficient comes from the anomalous thresholds!

\section{Integral formula for discontinuities}
\label{sec:Landau}

In the previous section, we've seen that the anomalous threshold can be isolated in a dispersive representation by identifying it as the double discontinuity of the amplitude for the same variable. In this section, we derive this in a more general setup to get more insights into the double-discontinuity formula.

As we've seen the anomalous threshold enters the physical sheet when one of the external kinematics become unstable. At this point the two roots of the triangle singularity becomes degenerate. Such kinematic configuration can be analyzed utilizing Landau equations.  We start with the $n$-point one-loop diagram
\begin{align*}
\begin{tikzpicture}
\draw (270:0.8) -- (330: 0.8) -- (30: 0.8) -- (90:0.8) -- (150: 0.8) --(210: 0.8);
\draw[dotted]  (210: 0.8) -- (270: 0.8);
\draw (90:0.8) -- (90:1.5) ;
\draw (30:0.8) -- (30:1.5) ;
\draw (330:0.8) -- (330:1.5) ;
\draw (150:0.8) -- (150:1.5) ;
\draw[dotted] (210:0.8) -- (210:1.5) ;
\draw[dotted] (270:0.8) -- (270:1.5) ;
\node at (90:1.8)  {$p_1$};
\node at (30:1.8)  {$p_2$};
\node at (330:1.8)  {$p_3$};
\node at (150:1.8)  {$p_n$};
\node at (60: 1) {$m_1$};
\node at (0: 1) {$m_2$};
\node at (120: 1) {$m_n$};
\end{tikzpicture}
\end{align*}
which is proportional to
\begin{align}
\mathcal{I}_n= \int \frac{\D^d \ell}{i \pi^{d/2}} \prod_{i=1}^n \frac{-1}{q_i^2+m_i^2}=\Gamma(n{-}\frac{d}{2}) \int^1_0 \left[ \prod_{i=1}^n \D \alpha_i \right]  \frac{\delta(1-\sum_i \alpha_i )}{[\frac{1}{2}\sum_{i,j} \alpha_i Y_{ij} \alpha_j]^{n-d/2}}
\label{n-point}
\end{align}
with $P_i=\sum_{j=1}^i p_i$ and
\begin{align}
Y_{ij}=-(P_i-P_j)^2-m_i^2-m_j^2
\,.
\end{align}
The final form of \eqref{n-point} manifests that $\mathcal{I}_n$ is a function of $z_{ij}=-(P_j-P_i)^2$ with $(j>i)$.

The singularities of $\mathcal{I}_n(z_{ij})$ are governed by Landau equations
\begin{align}
&\sum_{i,j} \alpha_i Y_{ij} \alpha_j=0
\,, \\
{\rm either}\quad &\alpha_i=0 \quad
{\rm or}\quad \sum_{j}  Y_{ij} \alpha_j=0
\quad \text{for each $i$}
\,.
\label{Landau_eq}
\end{align}
The singularities with $\alpha_i\neq 0$ for all $i$ are called leading singularities while those with $\alpha_i=0$ are called lower-order singularities. The lower-order singularities are the leading singularities for the diagram where lines $i$ corresponding to $\alpha_i=0$ are contracted.
Let $|{\bm Y}_{i_1 i_2 \cdots i_p}|,~(0\leq p\leq n-1)$ be the principal minor of order $p$ of ${\bm Y}=(Y_{ij})$ with removing the $i_1$th, $i_2$th, $\cdots$, and $i_p$th rows and columns. The equation
\begin{align}
|{\bm Y}_{i_1 i_2 \cdots i_p}|=0
\label{minorY}
\end{align}
solves the Landau equation \eqref{Landau_eq} with $\alpha_{i_1}=\alpha_{i_2}=\alpha_{i_p}=0$. Hence, \eqref{minorY} determine would-be singularity hypersurfaces in the complex $z_{ij}$-space which we call Landau surfaces. The singularities determined by \eqref{minorY} are understood as the singularities where all internal lines except $q_{i_1}, q_{i_2}, \cdots, q_{i_p}$ are on-shell.
Note that not all parts of the Landau surfaces are necessarily singularities of $\mathcal{I}_n$. We shall particularly refer to surfaces/singularities with $p$ $\alpha_i$s zero as Landau surfaces/singularities of order $n{-}p$ corresponding to singularities where $(n{-}p)$-propagators are on-shell. Thus the leading order Landau surface corresponds to all propagators on-shell, which in fixed dimensions implies a non-trivial constraint on the external kinematics.

\begin{figure}[h]
\centering
\includegraphics[width=0.8\linewidth]{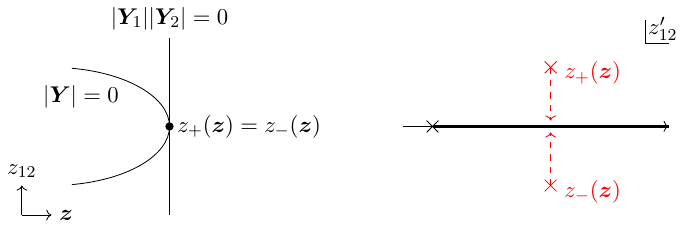}
\caption{Left: the roots of $|{\bm Y}|=0$, denoted by $z_{12}=z_{\pm}$ are multiple roots if and only if $|{\bm Y}_1|=0$ or $|{\bm Y}_2|=0$, showing these surfaces are tangent to each other. Right: the singularities of integrand $z_{12}'=z_{\pm}$ in the dispersive representation may pinch the integration contour $\mathcal{C}$ described by the thick line, generating a singularity of the integral at $z_+({\bm z})=z_-({\bm z})$. Note that $z_+=z_-^*$ if ${\bm z}, m_i^2 \in \mathbb{R}$ and $|{\bm Y}_1| | {\bm Y}_2| <0$ because $z_{\pm}$ are the roots of the quadratic equation \eqref{Jacobi_id2}.}
\label{fig:intersections}
\end{figure}

Let us consider intersections of two Landau surfaces. The equations of the Landau surfaces are all given by the principal minors. Hence, without loss of generality, we can consider intersections between the leading-order Landau surface ($n$-propagators on-shell) and a lower-order Landau surface. The equation $|{\bm Y}|=0$ is generically quadratic for one variable in $\{z_{ij} \}$. We bring this specified variable into the leading position without loss of generality, so we call it $z_{12}$. The set of variables other than $z_{12}$ is denoted by ${\bm z}$. The Laplace expansion of the determinant gives
\begin{align}
|{\bm Y}| = -z_{12}^2 |{\bm Y}_{12} | + \mathcal{O}(z_{12})
\,.
\end{align}
The coefficient of the highest degree term vanishes at the intersection with the Landau surface of order $n-2$, implying that one branch of the Landau surface of order $n$ reaches infinity at the intersection. Next, we consider the intersection between the Landau surfaces of order $n$ and order $n{-}1$ under the assumption $|{\bm Y}_{12}|\neq 0$ where the roots of $|{\bm Y}|=0$ are denoted by $z_{12}=z_{\pm}({\bm z})$. By Jacobi's theorem (for example, see \cite{aitken2017determinants,Eden:1966dnq}), we obtain the identity
\begin{align}
|{\bm Y}_1| |{\bm Y}_2| - |{\bm Y}^1_2|  |{\bm Y}^2_1| =|{\bm Y}| |{\bm Y}_{12}|
\,,
\label{Jacobi_id}
\end{align}
where $|{\bm Y}^i_j|$ denotes the $(i,j)$ algebraic minor of ${\bm Y}$. Note that $|{\bm Y}^i_j| = |{\bm Y}^j_i|$ because ${\bm Y}$ is symmetric. In particular, from eq.(\ref{Jacobi_id}) if $|{\bm Y}_{12}| \neq 0$ and $|{\bm Y}_1|=0$ or $|{\bm Y}_2|=0$, i.e. if we are considering the intersection with one lower-dimensional Landau surface, then equation $|{\bm Y}|=0$ is reduced to $|{\bm Y}^1_2|=0$. This implies that the different roots of $|{\bm Y}|=0$ are multiple roots $z_+=z_-$ at the intersection. In other words, a Landau surface is tangent to one lower-order Landau surface [see Fig.~\ref{fig:intersections} (left)]. The converse is also true: the equation $|{\bm Y}|=0$ is reduced to 
\begin{align}
|{\bm Y}_1| |{\bm Y}_2| - |{\bm Y}^1_2|^2 = 0
\,, \label{Jacobi_id2}
\end{align}
according to \eqref{Jacobi_id}. Since $|{\bm Y}_1|$ and $|{\bm Y}_2|$ are independent of $z_{12}$, while $|{\bm Y}^1_2|=z_{12} |{\bm Y}_{12}| + \mathcal{O}(z_{12}^0)$, the roots $z_{12}=z_{\pm}$ are degenerate only if $|{\bm Y}_1|=0$ or $|{\bm Y}_2|=0$. All in all, we conclude
\begin{align}
z_+({\bm z})=z_-({\bm z}) \iff |{\bm Y}_1| |{\bm Y}_2|=0
\,.
\label{zpm}
\end{align}
That is, when the order $n$ Landau singularity becomes degenerate, one in fact has an order $n{-}1$ Landau singularity. 

The relation \eqref{zpm} helps us to understand generations of singularities through the dispersive representation. Let us assume that the $n$-point one-loop function, with $n>2$, can be written as
\begin{align}
\mathcal{I}_n(z_{12}, {\bm z})= \int_{m_{\rm th}^2}^{\infty} \frac{\D z_{12}'}{2\pi i} \frac{\Disc_{z_{12}'} \mathcal{I}_n (z_{12}' , {\bm z})}{z_{12}'-z_{12}}
\,,
\label{In_dis}
\end{align}
for a certain region of ${\bm z}$.\footnote{Here, we consider the dispersive representation with zero subtraction, which should hold for triangle, box, and any higher $n$-gon in four dimensions. If subtractions are needed, a similar discussion can be made by replacing \eqref{In_dis} with a dispersive representation of derivatives of $\mathcal{I}_n$.} Here we treat the external kinematic variables to be independent and below two-particle normal thresholds. 
Suppose that $\Disc_{z_{12}} \mathcal{I}_n$ has a  singularity that can be identified as a Landau singularity of order $p$ at $z_{12}'=z_{\pm}({\bm z})$.  The $n$-point function $\mathcal{I}_n$ may possess the following types of pinch singularities: (i) one of $z_{\pm}$ coincides with the singularity coming from the denominator $z_{12}'=z_{12}$, trapping the integration contour (see Fig.~\ref{fig:generation}) and (ii) the singularities $z_{12}'=z_{\pm}$ pinch the integration contour [see Fig.~\ref{fig:intersections} (right)]. They occur at $z_{12}=z_{\pm}({\bm z})$ for the first case and at $z_+({\bm z})=z_-({\bm z})$ for the second case, respectively. The first, which is $z_{12}$-dependent is the condition for the Landau surface of order $p$ while the second, whose position is independent of $z_{12}$, is the condition for the surface of order $p-1$.  Said in another way, for the $z_{12}$ dispersive representation, $z_{12}$ independent singularities arise from lower order Landau surfaces.\footnote{Note that this is sufficient for the generations of singularities of $\mathcal{I}_n$ but not necessary. The function $\mathcal{I}_n$ may also be singular due to a pinch between Landau surfaces of different orders or an end-point singularity.}

We can derive an integral formula for the discontinuity by using the above discussion. Here, in addition to the variable $z_{12}$, we only vary an additional variable denoted as $z$. The other variables are supposed to be fixed so that the $n$-point function $\mathcal{I}_n$ is real analytic in the variables of interest $\{z_{12},z\}$. Let $z=z_s\in \mathbb{R}$ be the singularity of order $p{-}1$ of $\mathcal{I}_n(z_{12}, z)$ and let the dispersion relation
\begin{align}
\mathcal{I}_n(z_{12}, z)= \int_{m_{\rm th}^2}^{\infty} \frac{\D z_{12}'}{2\pi i} \frac{\Disc_{z_{12}'} \mathcal{I}_n (z_{12}' , z)}{z_{12}'-z_{12}}
\end{align}
holds in $z<z_s$ corresponding to $|{\bm Y}_1| |{\bm Y}_2|<0$. We consider the analytic continuation into the region $z>z_s$ corresponding to $|{\bm Y}_1| |{\bm Y}_2|>0$. As discussed, a singularity of order $p-1$ of $\mathcal{I}_n$ in the $z$-plane is generated by the pinch of a pair of singularities of order $p$ of $\Disc_{z_{12}'} \mathcal{I}_n$ in the $z_{12}'$-plane. We need to add a small imaginary part $\pm i \varepsilon$ to $z$ to avoid the pinch and shift the positions of singularities $z_{12}=z_{\pm}(z)$ away from the real axis in $z>z_s$.
Depending on the sign of the imaginary part, the positions of the singularities $z_{12}'=z_{\pm}$ will be different. 
According to the real analyticity, the singularities are found in the complex conjugate positions for the opposite sign of the imaginary part, $z_{\pm}(z+i\varepsilon)=[z_{\mp}(z-i\varepsilon)]^*$, implying that
\begin{align}
  \int_{m_{\rm th}^2}^{\infty}  \frac{\D z_{12}'}{2\pi i} \frac{\Disc_{z_{12}'} \mathcal{I}_n (z_{12}' , z+i\varepsilon)}{z_{12}'-z_{12}}
&=\adjincludegraphics[valign=c, scale=0.5]{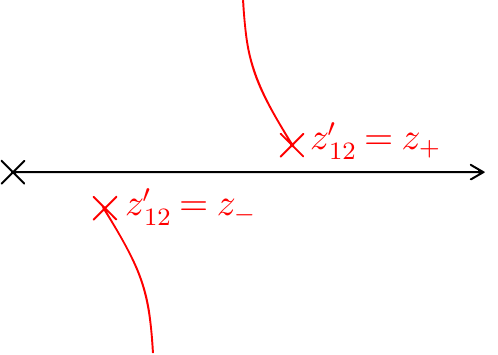}
\label{In_minus1}\\ \int_{m_{\rm th}^2}^{\infty}   \frac{\D z_{12}'}{2\pi i} \frac{\Disc_{z_{12}'} \mathcal{I}_n (z_{12}' , z-i\varepsilon)}{z_{12}'-z_{12}}
&= \adjincludegraphics[valign=c, scale=0.5]{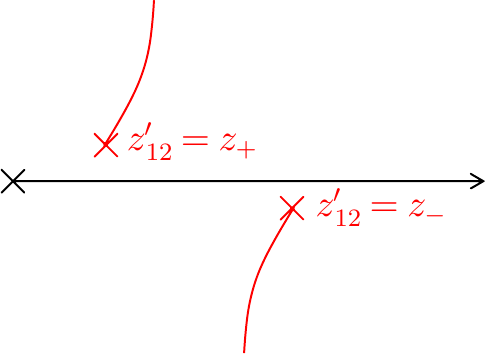}
\label{In_minus2}
\end{align}
where RHS represents the integration contour and the singularities of the integrand. The singularities $z_{12}'=z_{\pm}$ can be branch points of the integrated answer so we introduce branch cuts denoted by red curves.\footnote{If they are poles, the discussion is straightforward.} They should also be symmetrical for the opposite sign of imaginary parts $z\pm i\varepsilon$. Note that $z_+$ is the singularity above the integration contour before the analytic continuation $(z<z_s)$ and it remains above it in $z>z_s$. We further analytically continue the first equation into the lower-half $z$-plane, which is the same problem as we have done in Sec.~\ref{sec:dispersion_relation}:
\begin{align}
\adjincludegraphics[valign=c, scale=0.5]{contour1v2}
&\to 
\adjincludegraphics[valign=c, scale=0.5]{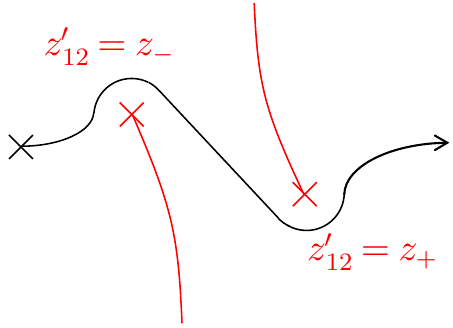}
\nonumber \\
&=
\adjincludegraphics[valign=c, scale=0.5]{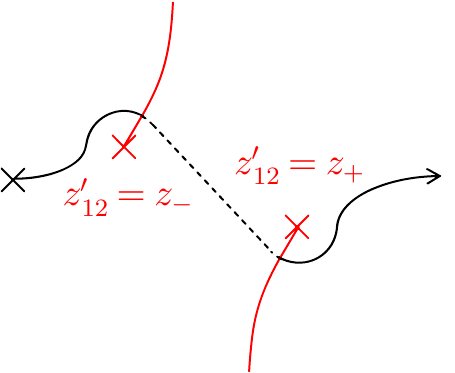} 
\nonumber \\
&=
\adjincludegraphics[valign=c, scale=0.5]{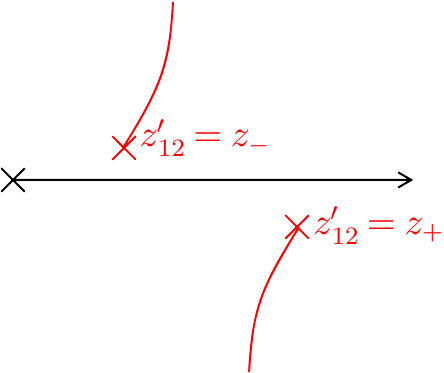} + \adjincludegraphics[valign=c, scale=0.5]{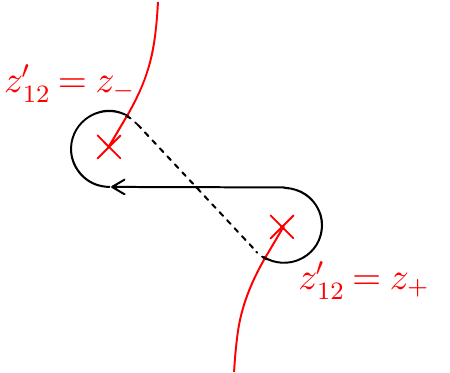}
\,. \label{deformation}
\end{align}
As in Sec.~\ref{sec:dispersion_relation}, the branch cuts are deformed so that the direction of the cuts agree with those of \eqref{In_minus2} . Then the contour integral is split into the real axis and the closed contour.
The first term on the third line is exactly the same as \eqref{In_minus2} after relabelling $z_{\pm}$ and the second term is replaced with the integral of the double discontinuity of the $z_{12}$-channel. Thus taking the difference between \eqref{In_minus1} and \eqref{In_minus2}, we obtain a new formula for discontinuities:
\begin{align}
\Disc_z \mathcal{I}_n(z_{12}, z) = 
\int_{z_+}^{z_-}  \frac{\D z_{12}'}{2\pi i} \frac{\Disc_{z_{12}'}^2 \mathcal{I}_n (z_{12}' , z)}{z_{12}'-z_{12}}
\,.
\label{discIn2}
\end{align}
The LHS is the discontinuity associated with the singularity of order $p-1$ in the $z$-plane whereas the RHS comes from the singularities of order $p$ in the $z_{12}$-plane. The information about different singularities in different channels is related through \eqref{discIn2}.

A key feature of \eqref{discIn2} is to relate singularities associated with different numbers of cuts. To illustrate it, let us consider the already-accustomed triangle diagram $(p=3)$. We apply the two-particle cut to the LHS and the three-particle cut to the RHS:
\begin{align}
{\rm LHS}=
\begin{tikzpicture}[baseline=-2]
\begin{feynhand}
\propag (0.8,0) --  (0, 0)  ;
\propag (-2, -1)  -- (-1.2, -0.6);
\propag[ultra thick] (-1.2, -0.6) -- (0, 0) ;
\propag (-2, 1) -- (-1.2 , 0.6) ;
\propag (-1.2, 0.6) -- (0, 0);
\propag (-1.2,-0.6) -- (-1.2,0.6)  ;
\draw[dashed, red] (-1.5, 0) -- (-0.9,0);
\draw[dashed, red] (-0.7, 0.1) -- (-0.4,0.6);
\end{feynhand}
\end{tikzpicture}
\,, \qquad \qquad 
{\rm RHS}=
\begin{tikzpicture}[baseline=-2]
\begin{feynhand}
\propag (0.8,0) --  (0, 0)  ;
\propag (-2, -1)  -- (-1.2, -0.6);
\propag[ultra thick] (-1.2, -0.6) -- (0, 0) ;
\propag (-2, 1) -- (-1.2 , 0.6) ;
\propag (-1.2, 0.6) -- (0, 0);
\propag (-1.2,-0.6) -- (-1.2,0.6)  ;
\draw[dashed, red] (-1.5, 0) -- (-0.9,0);
\draw[dashed, red] (-0.7, 0.1) -- (-0.4,0.6);
\draw[dashed, red] (-0.7, -0.1) -- (-0.4,-0.6);
\end{feynhand}
\end{tikzpicture}
\,.
\end{align}
Suppose that the thin internal lines correspond to IR states and the thick line is a UV state. Then, the two-particle cut is possible at IR while the three particles can be on-shell only at UV. In this way, the formula \eqref{discIn2} can connect singularities at different energies through the integration. This is precisely what we have seen in \eqref{B2dispersionA}, the triangle cut arises from the double discontinuity in $z_{12}$. The UV double discontinuity arises because we have crossed the IR branch cut of the external mass variable. Further checks of \eqref{discIn2} for $n=3,4$ are given in Appendix~\ref{sec:check}.

\section{Conclusions}
\label{sec:conclusion}
Unstable particles have been largely unexplored in studies of S-matrix. The obvious difficulty comes from the fact that unstable particles do not appear in the asymptotic states so the definition of scattering amplitudes is obscure. Precisely, scattering amplitudes of unstable particles can be defined by the analytic continuation of higher-point amplitudes into an unphysical sheet where the pole associated with the unstable particle exists. The difficulty then traces back to the lack of precise knowledge of amplitudes away from the physical sheet.

In this work, we have studied the analytic properties of scattering amplitudes in unstable kinematics. The main results are twofold: there exist anomalous thresholds in a {\it UV region} of the energy variable and such anomalous thresholds can yield {\it negative} contribution to the dispersion relation. Both properties are quite different from what we have learned from stable particles. The appearance of anomalous thresholds in scatterings of heavy particles has been well known (see e.g.~\cite{Hannesdottir:2022bmo, Correia:2022dcu} for recent papers). When an external mass is analytically continued to a heavy value, singularities which are originally situated on the second sheet, enter the first sheet through the IR side of the branch cut. These IR singularities associated with light loops appear in both stable and unstable kinematics and can be traced by low-energy EFTs. On the other hand, if the external mass is analytically continued beyond the decay threshold, singularities associated with loops involving particle(s) whose mass is much heavier than the external masses also enter the first sheet through the UV side of the branch cut. The UV singularities give rise to an unsubtractable ``anomalous'' contribution to the dispersion relation in the form of double discontinuity. At least in the example, we have found that this anomalous contribution is negative, violating the positivity bounds known in stable-particle scatterings. 

Our results can be phrased in another way. For the scattering of stable particles, the anomalous threshold can only enter to the physical sheet by passing through the physical threshold at $4m^2$, and thus strictly within the realm of IR kinematics. In defining our unstable particle S-matrix by analytically continuing through the decay threshold to a complex mass, we introduce a new avenue where the anomalous threshold can enter the physical sheet. In such case, ``UV'' anomalous thresholds, which for us are triangle singularities associated with UV states, can also enter the physical sheet. Note that not all is lost. As seen in our toy model analysis, the leading contribution to $B_2$ comes from $\mathcal{I}_{\rm small}$ in eq.~(\ref{divide_box}), i.e. where the loop momentum is small compared to $m^2_H$. In this region, the four-point amplitude $\Amp(\pi\pi \chi_L\chi_L)$ is well approximated by an EFT description with local operators suppressed by $m^2_H$ and $B_2$ is calculable by EFT loops with such local vertices. Thus such UV anomalous threshold is well computable in terms of a handful of EFT Wilson coefficients. It will be interesting to systematically study such effects in the future. 

At first glance this conclusion might seem contradictory. On the one-hand the anomalous threshold is of ``UV'' origin as it is associated with a triangle singularity that has a UV massive propagator. On the other, in the explicit computation, the leading contributions came from $\mathcal{I}_{\rm small}$ in eq.~(\ref{divide_box}), which is computable directly using low energy EFT, i.e. in the IR. The resolution is of course \eqref{discIn2}, where the discontinuity of the variable $z$ in the IR (LHS), knows about the singularities in the UV in the variable $z_{12}$ (RHS). The essential ingredients for the derivation of the formula \eqref{discIn2} are real analyticity, the dispersion relation, and its analytic continuation. The analysis of Landau equations that led to the relations between different singularities \eqref{zpm} are specialized to one-loop, while other parts of the discussion, e.g. the contour deformation \eqref{deformation}, are expected to be generic. Thus if we have control over the position of the singularities, similar formulae may be derived beyond the one-loop level or even non-perturbatively by starting with the (twice-subtracted) dispersion relation. The precise analysis is left for future investigation. Also, the formula is applicable not only for unstable particles but also for stable-particle scatterings. In this case, one can think of the LHS as the EFT-calculable $t$-channel discontinuity of the 2-to-2 amplitude. The RHS corresponds to the $s$- and $u$-channel double discontinuity associated with one cut added to the LHS, which can be a UV state. One such example is a box integral like \eqref{M_example} where it can have an IR $t$-channel triangle singularity and a UV $s$-channel box singularity. We can then obtain a new type of UV-IR relation of the 2-to-2 scattering amplitude.
However, we need more knowledge about the anomalous thresholds and the double discontinuity to understand such a UV-IR relation, which we leave for future work.

In summary, our results pose new challenges for the S-matrix bootstrap. On the one hand, our analysis indicates that the standard dispersive formulae cannot be immediately applied to unstable particles except in the absence of anomalous thresholds, e.g.~the tree-level approximation, or assuming a model that the interaction is dominated by a non-decay process. Since most particles have finite decay width, to be agnostic about models, one needs to find an appropriate prescription for phenomenological applications such as the standard model effective field theory and strongly coupled systems. On the other hand, it is evident that our knowledge is only the tip of the iceberg of the S-matrix. We have found that singularities of different channels at different energies are related, which would be a portion of the hidden structure of the S-matrix. Scattering amplitudes should be even more restricted than what we currently know and studying unstable particles, or more generically speaking, singularities other than normal thresholds (resonance poles, antibound state poles, anomalous thresholds, etc) will help us expose the entire structure of the S-matrix.

\section*{Acknowledgements}
We would like to thank Holmfridur Hannesdottir and Sebastian Mizera for discussions and comments on the draft. The Feynman diagrams in this paper were drawn with the help of \texttt{TikZ-FeynHand}\, \cite{Ellis:2016jkw}. K.A. is grateful to the organizers of the workshop ``14th Taiwan String Workshop'' and the hospitality of NTU where part of this work was carried out. Y-t H would like to thank the hospitality and support of YITP, during which a majority of this work was completed. The work of K.A. was supported by JSPS Grants-in-Aid for Scientific Research, No.~20K14468 and No.~24K17046. The work of Y-t H is supported by NSTC grant no. 112-2811-M-002 -054 -MY2.

\appendix
\section{Explicit check of discontinuity formula}
\label{sec:check}
In this appendix, we compute the LHS and RHS of \eqref{discIn2} and confirm the agreement. We focus on the triangle and box diagrams in four dimensions. The explicit forms of the discontinuities are given in \eqref{rho_tri} for $s_1$ and \eqref{rho_box} for $s_{12}$, which are worth rewriting here as a reference:
\begin{align*}
\rho_{\rm tri}&=\frac{1}{\lambda^{1/2}(s_1,s_2,s_3)} 
\ln\left[ \frac{ \sqrt{ 2s_1 S_{\rm tri} + \lambda(s_1, m_2^2, m_3^2) \lambda(s_1, s_2, s_3)  }- \sqrt{\lambda(s_1, m_2^2, m_3^2) \lambda(s_1,s_2,s_3)} }
{ \sqrt{2s_1 S_{\rm tri} + \lambda(s_1, m_2^2, m_3^2) \lambda(s_1, s_2, s_3) } + \sqrt{\lambda(s_1, m_2^2, m_3^2) \lambda(s_1,s_2,s_3)}   }
\right]
\tag{\ref{rho_tri}}
\\
\rho_{\rm box}&=\frac{1}{S_{\rm box}^{1/2}} 
\ln\left[ \frac{ \sqrt{S_{\rm tri}^L S_{\rm tri}^R+\lambda(s_{12}, m_2^2,m_4^2) S_{\rm box} } + \sqrt{\lambda(s_{12}, m_2^2, m_4^2) S_{\rm box}} }
{ \sqrt{S_{\rm tri}^L S_{\rm tri}^R+\lambda(s_{12}, m_2^2,m_4^2) S_{\rm box} } - \sqrt{\lambda(s_{12}, m_2^2, m_4^2) S_{\rm box}}  }
\right]
. \tag{\ref{rho_box}}
\end{align*}

\begin{figure}[t]
\centering
 \includegraphics[width=0.45\linewidth]{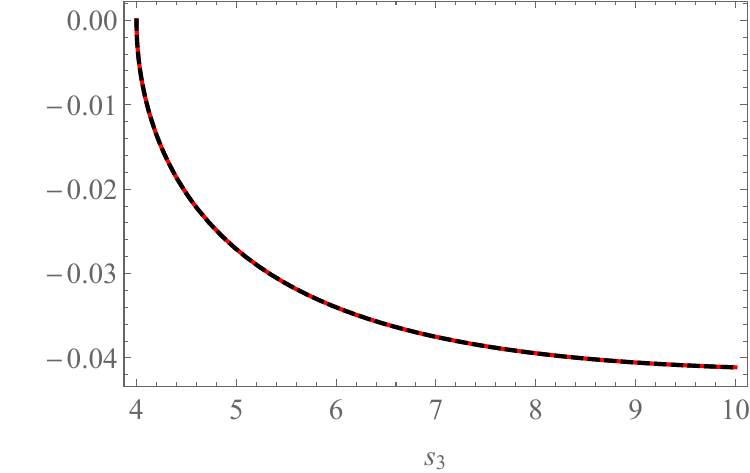}	
 \quad
  \includegraphics[width=0.45\linewidth]{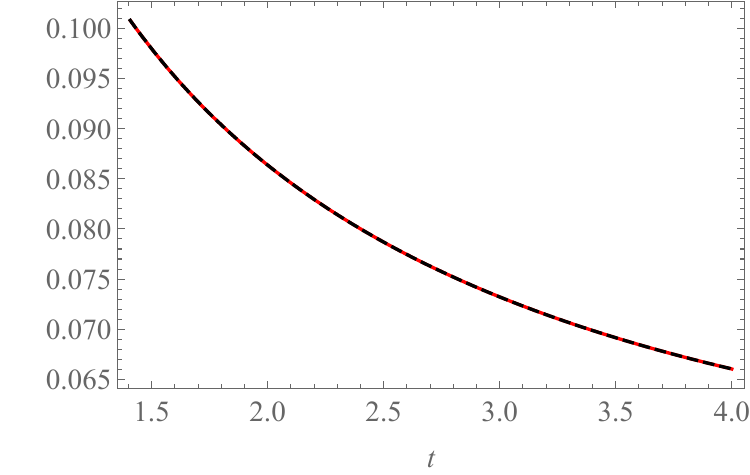}	
\caption{Comparisons between the LHS and the RHS of \eqref{discI3} (left) and \eqref{discI4} (right). The red curves and the black dashed curves respectively represent the LHS and the RHS divided by $2\pi i$. Here, we set $s_1=s_2=m_1^2=m_2^2=1, m_3=4$ (left) and $s=m_L^2=1, M=1.9, m_H=4$ (right).}
\label{fig:check}
\end{figure}

We first consider the triangle diagram and see the relation between the two-particle and three-particle cuts:
\begin{align}
\begin{tikzpicture}[baseline=-2]
\begin{feynhand}
\propag (0.8,0) --  (0, 0)  ;
\propag (-2, -1)  -- (-1.2, -0.6);
\propag[ultra thick] (-1.2, -0.6) -- (0, 0) ;
\propag (-2, 1) -- (-1.2 , 0.6) ;
\propag (-1.2, 0.6) -- (0, 0);
\propag (-1.2,-0.6) -- (-1.2,0.6)  ;
\draw[dashed, red] (-1.5, 0) -- (-0.9,0);
\draw[dashed, red] (-0.7, 0.1) -- (-0.4,0.6);
\node at (-2.2,1) {$3$};
\node at (-2.2,-1) {$2$};
\node at (1,0) {$1$};
\end{feynhand}
\end{tikzpicture}
&=\Disc_{s_3}\mathcal{I}_{\rm tri}=2\pi i \rho_{\rm tri}|_{1\leftrightarrow 3}
\,, \\
\begin{tikzpicture}[baseline=-2]
\begin{feynhand}
\propag (0.8,0) --  (0, 0)  ;
\propag (-2, -1)  -- (-1.2, -0.6);
\propag[ultra thick] (-1.2, -0.6) -- (0, 0) ;
\propag (-2, 1) -- (-1.2 , 0.6) ;
\propag (-1.2, 0.6) -- (0, 0);
\propag (-1.2,-0.6) -- (-1.2,0.6)  ;
\draw[dashed, red] (-1.5, 0) -- (-0.9,0);
\draw[dashed, red] (-0.7, 0.1) -- (-0.4,0.6);
\draw[dashed, red] (-0.7, -0.1) -- (-0.4,-0.6);
\node at (-2.2,1) {$3$};
\node at (-2.2,-1) {$2$};
\node at (1,0) {$1$};
\end{feynhand}
\end{tikzpicture}
&=\Disc_{s_1}^2\mathcal{I}_{\rm tri}=
\frac{(2\pi i)^2}{\lambda^{1/2}(s_1,s_2,s_3)}
\,,
\end{align}
where the thick line is supposed to be heavy so that the triangle singularities appear in the first sheet of the $s_1$ plane (See Sec.~\ref{sec:singularityUV}). The (single) discontinuity in $s_3$ is given by \eqref{rho_tri} with the replacement $1\leftrightarrow 3$.
The double discontinuity is computed from the discontinuity \eqref{rho_tri} for the logarithmic singularity $S_{\rm tri}=0$. Then, the formula \eqref{discIn2} yields
\begin{align}
    2\pi i \rho_{\rm tri}|_{1\leftrightarrow 3}
    =\int_{s_+}^{s_-} \frac{\D s_1'}{2\pi i} \frac{(2\pi i)^2}{\lambda^{1/2}(s_1',s_2,s_3)} \frac{1}{s_1'-s_1}
    \,, \label{discI3}
\end{align}
where $s_1=s_{\pm}$ are the roots of $S_{\rm tri}=0$, which are explicitly given by \eqref{s1roots}. Here, we should make sure that only the triangle singularities $S_{\rm tri}=0$ deform the contour, in particular, $\lambda(s_1',s_2,s_3)>0$ in the interval of integration. As demonstrated in Fig.~\ref{fig:check}, one can confirm that the equality \eqref{discI3} indeed holds.

Next, we discuss the box integral $\mathcal{I}_{\rm box}$ to relate the $s$ and $t$ channel singularities. For a sufficiently large external mass (which can be stable), the triangle singularity appears below the normal threshold. In such a case, the analytic continuation of the $s$-channel dispersion relation in $t$ first touches the triangle singularity. We can thus apply \eqref{discIn2} to the three- and four-particle cuts of the box diagram:
\begin{align}
{\rm LHS}=
\begin{tikzpicture}[baseline=0]
\begin{feynhand}
\propag (-1.2, -0.8) node [left] {$1$} -- (-0.6, -0.6) ;
\propag (-1.2, 0.8) node [left] {$2$} -- (-0.6, 0.6) ;
\propag (1.2, 0.8) node [right] {$3$} -- (0.6, 0.6) ;
\propag (1.2, -0.8) node [right] {$4$} -- (0.6, -0.6) ;
\propag (-0.6, -0.6) -- (-0.6, 0.6) node [midway, left=0.1] {$m_L$};
\propag (-0.6, 0.6) -- (0.6, 0.6) node [midway, above=0.1] {$m_L$};
\propag (0.6, 0.6) -- (0.6, -0.6) node [midway, right=0.1] {$m_L$};
\propag[ultra thick] (0.6, -0.6) -- (-0.6, -0.6) node [midway, above=0.1] {$m_H$};
\draw[dashed, red] (-0.8, 0) -- (-0.4,0);
\draw[dashed, red] (0.8, 0) -- (0.4,0);
\draw[dashed, red] (0, 0.8) -- (0, 0.4);
\end{feynhand}
\end{tikzpicture}
\,, \qquad
{\rm RHS}=
\begin{tikzpicture}[baseline=0]
\begin{feynhand}
\propag (-1.2, -0.8) node [left] {$1$} -- (-0.6, -0.6) ;
\propag (-1.2, 0.8) node [left] {$2$} -- (-0.6, 0.6) ;
\propag (1.2, 0.8) node [right] {$3$} -- (0.6, 0.6) ;
\propag (1.2, -0.8) node [right] {$4$} -- (0.6, -0.6) ;
\propag (-0.6, -0.6) -- (-0.6, 0.6) node [midway, left=0.1] {$m_L$};
\propag (-0.6, 0.6) -- (0.6, 0.6) node [midway, above=0.1] {$m_L$};
\propag (0.6, 0.6) -- (0.6, -0.6) node [midway, right=0.1] {$m_L$};
\propag[ultra thick] (0.6, -0.6) -- (-0.6, -0.6) node [midway, above=0.1] {$m_H$};
\draw[dashed, red] (-0.8, 0) -- (-0.4,0);
\draw[dashed, red] (0.8, 0) -- (0.4,0);
\draw[dashed, red] (0, 0.8) -- (0, 0.4);
\draw[dashed, red] (0, -0.8) -- (0, -0.4);
\end{feynhand}
\end{tikzpicture}
\,.
\end{align}
For simplicity, we assume the same external masses $p_i^2=-M^2$ and the internal masses specified by the diagrams. Note that \eqref{rho_box} is the discontinuity across the normal threshold cut. It cannot be directly applied to evaluate the discontinuity across the triangle cut. However, we can find the triangle discontinuity as follows. In $2m_L^2<t<4m_L^2$, with a small fixed $s$, the $t$-channel dispersion relation should take the form
\begin{align}
    \mathcal{I}_{\rm box}&=\int_{t_{\rm tri}}^{\infty} \frac{\D t'}{2\pi i} \frac{\Disc_t \mathcal{I}_{\rm box}}{t'-t}
    \nonumber \\
    &= \left( \int_{t_{\rm tri}}^{4m_L^2} +\int_{4m_L^2}^{\infty} \right)\frac{\D t'}{2\pi i} \frac{\Disc_t \mathcal{I}_{\rm box}}{t'-t}
    \,,
    \label{disp_box_t}
\end{align}
with $t_{\rm tri}=4M^2-M^4/m_L^2$. This dispersion relation should agree with the analytic continuation of \eqref{disp_box} in the external mass after appropriately relabelling variables. As shown in Fig.~\ref{fig:generation} (right), the analytic continuation generates the additional integral below the normal threshold, which should be identified with the first integral of the second line of \eqref{disp_box_t}. The contour integral of Fig.~\ref{fig:generation} (right) is nothing but the integral of the discontinuity of $\rho_{\rm box}$ for the triangle singularity which is logarithmic. Hence, the triangle cut is
\begin{align}
    \Disc_{t} \mathcal{I}_{\rm box} & = \frac{(2\pi i)^2}{S_{\rm box}^{1/2}}
     \qquad (t_{\rm tri}<t<4m_L^2) \,.
     \label{disct_tri}
\end{align}
On the other hand, the $s$-channel double discontinuity is computed by the discontinuity \eqref{rho_box} for the box singularity $S_{\rm box}=0$. The singularity arises from the denominator of \eqref{rho_box}. Note that $S_{\rm box}=0$ is a singularity of $\rho_{\rm box}$ only if the logarithm does not take a principal value. Therefore, the double discontinuity should be\footnote{To determine the overall sign, we need a precise analysis of the analytic continuation. We, however, simply determine it to require the agreement with \eqref{disct_tri} through \eqref{discIn2}.}
\begin{align}
    \Disc^2_{s} \mathcal{I}_{\rm box} &= -2\times \frac{(2\pi i)^2}{S_{\rm box}^{1/2}}
    \,.
\end{align}
The integral formula \eqref{discIn2} then gives
\begin{align}
    \frac{(2\pi i)^2}{S_{\rm box}^{1/2}}
    =\int_{s^b_+}^{s^b_-} \frac{\D s'}{2\pi i} \frac{(-2) \times (2\pi i)^2}{S_{\rm box}^{1/2}} \frac{1}{s'-s}
    \,,
    \label{discI4}
\end{align}
where $s=s_{\pm}^b$ are the roots of $S_{\rm box}=0$, satisfying $s_-^b<s_+^b$. One can confirm the agreement as shown in Fig.~\ref{fig:check}.

One can notice that \eqref{discI3} and \eqref{discI4} are the same as the dispersion relations of the discontinuities. For instance, $\Disc_t \mathcal{I}_{\rm box}=(2\pi i)^2/S_{\rm box}^{1/2}$ has a branch cut in $s_-^b<s<s_+^b$ in the complex $s$ plane, corresponding to the change of the sign of $S_{\rm box}$. Hence, we obtain
\begin{align}
    \Disc_t \mathcal{I}_{\rm box} &= \int_{s_-^b}^{s_+^b}\frac{\D s'}{2\pi i} \frac{\Disc_s \Disc_t \mathcal{I}_{\rm box}}{s'-s}
    \nonumber \\
    &=\int_{s_-^b}^{s_+^b} \frac{\D s'}{2\pi i} \frac{2 \times (2\pi i)^2}{S_{\rm box}^{1/2}} \frac{1}{s'-s}
\end{align}
which exactly agrees with \eqref{discI4}. However, we emphasise that the integrand of \eqref{discI4} is the double discontinuity in the same variable, rather than $s$ and $t$, because it is derived by the analytic continuation of the $s$-channel dispersion relation. In other words, the double discontinuities are the same
\begin{align}
\Disc_s \Disc_t \mathcal{I}_{\rm box}
=-\Disc_s^2 \mathcal{I}_{\rm box}
\end{align}
for the box singularity $S_{\rm box}=0$.
Here, we recall that $\Disc_s^2$ is defined by (the first sheet value of $\Disc_s$) $-$ (the second sheet value of $\Disc_s$) while $\Disc_s \Disc_t =$ (the upper-half plane of $\Disc_t$) $-$ (the lower-half plane of $\Disc_t$).

\bibliography{ref}

\end{document}